\documentclass[superscriptaddress,reprint,amsmath,amssymb,floatfix, aps, prx]{revtex4-1}

\usepackage{xcolor} 
\usepackage{graphicx}
\usepackage{dcolumn}
\usepackage{upgreek}
\usepackage{bm}
\usepackage{hyperref}
\usepackage{floatrow}
\usepackage{times}
\usepackage[mathlines]{lineno}
\usepackage{physics}

\begin{document}

\title{Experimental Realization and Characterization of Stabilized Pair Coherent States}
\author{Jeffrey M.~Gertler$^\dagger$}
\affiliation{Department of Physics, University of Massachusetts-Amherst, Amherst, MA, 01003, USA}
\author{Sean van Geldern$^\dagger$}
\affiliation{Department of Physics, University of Massachusetts-Amherst, Amherst, MA, 01003, USA}
\author{Shruti Shirol}
\affiliation{Department of Physics, University of Massachusetts-Amherst, Amherst, MA, 01003, USA}
\author{Liang Jiang}
\affiliation{Pritzker School of Molecular Engineering, University of Chicago, Chicago, IL, USA}
\author{Chen Wang$^*$}
\affiliation{Department of Physics, University of Massachusetts-Amherst, Amherst, MA, 01003, USA}
\email{wangc@umass.edu}

\date{\today}

\def\thefootnote{*}\footnotetext{These authors contributed equally to this work}\def\thefootnote{\arabic{footnote}}

\begin{abstract}
The pair coherent state (PCS) is a theoretical extension of the Glauber coherent state to two harmonic oscillators.  It is an interesting class of non-Gaussian continuous-variable entangled state and is also at the heart of a promising quantum error correction code: the pair cat code. 
Here we report an experimental demonstration of the pair coherent state of microwave photons in two superconducting cavities.  We implement a cross-cavity pair-photon driven dissipation process, which conserves the photon number difference between cavities and stabilizes the state to a specific complex amplitude. We further introduce a technique of quantum subspace tomography, which enables direct measurements of individual coherence elements of a high-dimensional quantum state without global tomographic reconstruction.  We characterize our two-mode quantum state with up to 4 photons in each cavity using this subspace tomography together with direct measurements of the photon number difference and the joint Wigner function.   We identify the spurious cross-Kerr interaction between the cavities and our dissipative reservoir mode as a prominent dephasing channel that limits the steady-state coherence in our current scheme. Our experiment provides a set of reservoir engineering and state characterization tools to study quantum optics and implement multi-mode bosonic codes in superconducting circuits.  
\end{abstract}

\maketitle

\section{Introduction}
The use of continuous-variable states of bosonic modes as a platform for quantum information processing, originating in quantum optics~\cite{dodonov_nonclassical_2002, braunstein_quantum_2005}, is rapidly advancing in superconducting circuit quantum electrodynamics (QED)~\cite{joshi_quantum_2021, ma_quantum_2021}. While many of the exotic states envisioned decades ago remain challenging to implement in the optical domain, they have become practical and valuable resources in the microwave domain thanks to the ability to engineer a wide range of mode couplings and nonlinearities in Josephson circuits~\cite{blais_circuit_2020}. For example, the Schr\"odinger cat states~\cite{vlastakis_deterministically_2013, wang_schrodinger_2016,grimm_stabilization_2020} and the Gottesman-Kitaev-Preskill grid states~\cite{campagne-ibarcq_quantum_2020} have not only been realized but also actively pursued for encoding logical qubits with error suppression or correction capabilities.

One interesting class of bosonic states yet to be studied experimentally is the pair coherent state (PCS), an example of a Barut-Girardello generalized coherent state~\cite{barut_new_1971}. This state gained theoretical interest as an example of a highly entangled two-mode state~\cite{agarwal_generation_1986, agarwal_Quantitative_2005} and was initially proposed to explain a suppression of amplified spontaneous emission in an atomic system~\cite{malcuit_suppression_1985}.  Analogous to the Glauber coherent state, $\ket{\alpha} = \mathcal{N} \sum_{n=0}^{\infty} \frac{\alpha^{n}}{\sqrt{n!}}\ket{n}$, a PCS can be written in the Fock state basis of two harmonic oscillators ($a$ and $b$) as:
\begin{equation}
    \ket{\gamma,\delta} = \mathcal{N} \sum_{n=0}^{\infty} \frac{\gamma^{n+\delta/2}}{\sqrt{n!(n+\delta)!}}\ket{n+\delta}_a\ket{n}_b
    \label{eq:PC_state}
\end{equation}
where $\delta$ is an integer describing the photon number difference (PND) between the two modes, $\gamma$ a complex number describing the amplitude and phase of the state, and $\mathcal{N}$ a normalization factor.  This state is both an eigenstate of the pair photon annihilation operator $\hat{a}\hat{b}$ and the PND operator $\hat{\delta}=\hat{a}^\dagger \hat{a} - \hat{b}^\dagger \hat{b}$:  
\begin{equation}
    \hat{a}\hat{b} \ket{\gamma,\delta} = \gamma \ket{\gamma,\delta}, 
    \hat{\delta}\ket{\gamma,\delta} = \delta\ket{\gamma,\delta}. 
\end{equation}
A PCS is inseparable and already in the form of a Schmidt decomposition~\cite{agarwal_Quantitative_2005}. The $\ket{\gamma, 0}$ state resembles a two-mode squeezed state in terms of photon number correlation~\cite{zhang_entanglement_2018} but has a Poisson-like photon number distribution approximately centered around $\gamma$. 

Pair coherent states form the basis of a recently-proposed quantum error correction (QEC) code called the pair cat code~\cite{albert_pair-cat_2019}. This code promises autonomous QEC of all types of first-order physical errors associated with loss \& dephasing by encoding a logical qubit in the superposition of pair coherent states $|\pm\gamma,\delta\rangle$ of two oscillators. 
The scheme involves stabilizing PND to correct the quantum jumps of photon loss events while simultaneously stabilizing the pair coherent state manifold with a two-mode four-photon dissipation process. 

Despite progress in universal control~\cite{ heeres_implementing_2017, gao_entanglement_2019, chakram_multimode_2022} and measurement feedback~\cite{ofek_extending_2016, campagne-ibarcq_quantum_2020, ma_error-transparent_2020, essig_multiplexed_2021} in bosonic cavities systems, there is a clear need for new tools for multi-mode quantum operations. For example, unitary preparation of a PCS using a dispersively coupled ancilla qubit with the current standard technique, numerical optimal control pulses~\cite{heeres_implementing_2017}, becomes prohibitively difficult for modest photon numbers. On the other hand, reservoir engineering~\cite{poyatos_quantum_1996} has been of particular interest for its ability to stabilize non-classical oscillator states in a resource-efficient manner~\cite{kienzler_quantum_2015}.  Moreover, engineered dissipation can provide us with not only encoded qubits with high noise bias~\cite{mirrahimi_dynamically_2014,albert_pair-cat_2019} but also a set of bias-preserving gates for hardware efficient quantum computing~\cite{guillaud_repetition_2019, puri_bias-preserving_2020, chamberland_building_2022, yuan_construction_2022}.  In circuit QED, realization of nonlinear dissipation operators have led to stabilization of the cat-state manifolds~\cite{leghtas_confining_2015, touzard_coherent_2018, lescanne_exponential_2020} and autonomous QEC of photon losses~\cite{gertler_protecting_2021} in a single cavity.  However, engineered nonlinear dissipation across two cavity modes remains to be explored.

\begin{table*}[tbp]
\caption{Comparison of multi-photon steady states under driven dissipative processes constructed with different operators, $\hat{o}=\hat{a},\hat{a}^2$ and $\hat{a}\hat{b}$.  For each of the three cases, we consider cavity 
dynamics following the master equation $\Dot{\rho}=-\frac{i}{\hbar}[\hat{H_0}+\hat{H_d},\rho]+\mathcal{D}(\rho)$ written in the rotating frame of the drives, where $\mathcal{D}$ is the Lindblad superoperator with $\hat{o}$ as its jump operator.  
In the textbook example of the driven damped quantum oscillator, the mode detuning $\Delta$ and single-photon loss $\kappa$ jointly counter the driving force and determine the complex amplitude of the unique steady state of the system.  
In two-photon dynamics, the complex amplitude of the steady-states is analogously determined by the one-mode or two-mode squeezing drives countered by the corresponding two-photon loss rates and confining Kerr Hamiltonian~\cite{gautier_combined_2022}. Our experiment produces PCS under both dissipative and cross-Kerr Hamiltonian confinement, with the effective cross-Kerr $K_\mathrm{eff}$ a few times stronger than the pair dissipation $\kappa_{ab}$. Notably, two steady states exist (with even or odd photon number parity $\Pi$) for the case of cat-state stabilization while there are infinitely many steady states (with different photon number difference $\delta$) for the case of stabilizing pair coherent states.  
}
\centering
\begin{tabular}{c c c c c}
\hline\hline\\[-2ex]
		Process category & Drive Hamiltonian $\hat{H}_d/\hbar$ & \,\,\, Dissipator $\mathcal{D}$ \,\,\, & Hamiltonian $\hat{H_0}/\hbar$ & Steady State(s) \\[0.5ex]
\hline\\[-1.5ex]
Single-photon & $\epsilon^*\hat{a}+\epsilon\hat{a}^\dagger$	& $\kappa\mathcal{D}[\hat{a}]$ &  $\Delta \hat{a}^\dagger\hat{a}$  & Coherent state $|\alpha\rangle$, $\alpha = \frac{\epsilon}{i\kappa/2-\Delta}$ \\[0.8ex]
Single-mode two-photon  & $\epsilon_2^*\hat{a}^2+\epsilon_2\hat{a}^{\dagger 2}$ ~\cite{slusher_observation_1985} 	& $\kappa_2\mathcal{D}[\hat{a}^2]$ \cite{leghtas_confining_2015} &  $K_{a} \hat{a}^{\dagger2}\hat{a}^2$ ~\cite{grimm_stabilization_2020}  & Cat states $|\alpha,\Pi=\pm1\rangle$, $\alpha = \sqrt{\frac{\epsilon_2}{i\kappa_2/2-K_{a}}}$\\[0.8ex]
Two-mode pair-photon  & $\epsilon_{ab}^*\hat{a}\hat{b}+\epsilon_{ab}\hat{a}^\dagger\hat{b}^\dagger$ ~\cite{heidmann_observation_1987}	& $\kappa_{ab}\mathcal{D}[\hat{a}\hat{b}]$ &  $K_\text{eff} \hat{a}^\dagger\hat{a}\hat{b}^\dagger\hat{b}$  & Pair coherent states $|\gamma, \delta\in\mathcal{Z}\rangle$, $\gamma = \frac{\epsilon_{ab}}{i\kappa_{ab}/2-K_\text{eff}}$\\[1ex]
\hline
\end{tabular}
\label{table:PC_principle}
\end{table*}

Similarly, characterization of multi-mode bosonic states poses substantial challenges beyond their single-mode counterparts.  Although multi-mode Wigner tomography using joint parity or joint photon number measurements has been previously demonstrated~\cite{wang_schrodinger_2016, chakram_multimode_2022, wang_deterministic_2011,  wollack_quantum_2022}, 
such process shares similar scalability challenges as in multi-qubit state tomography and additionally lacks the convenience to arrange orthogonal measurement basis for efficient information extraction.  
In order to explore the space of multi-mode bosonic QEC codes such as the pair-cat code or the two-mode binomial~\cite{chuang_bosonic_1997,michael_new_2016} and GKP codes~\cite{royer_encoding_2022}, it is crucial to develop efficient tools to characterize the relevant metrics of the states. 

In this work, we present an experimental realization of the pair coherent state and efficient characterization of its coherence.  
We expand the toolbox of quantum reservoir engineering by realizing an effective dissipation operator 
that removes photon pairs from two superconducting cavities and stabilizes the complex amplitude of the pair coherent state.  This pair-photon dissipation is realized using a pumped superconducting transmon ancilla for nonlinearity and a short-lived oscillator mode as the reservoir.  
To characterize the two-cavity state, we use three levels of the transmon ancilla to isolate selected subspaces of the large Hilbert space before Ramsey-style tomographic measurements.  We use this subspace tomography technique to independently measure the quantum coherence between individual pairs of Fock components  
of the two-cavity state. Our characterization reveals the phase distortion and limited coherence of the stabilized PC state, which we attribute to the spurious cavity-reservoir cross-Kerr interactions.  We further demonstrate an effective method of measuring PND without fine matching of system parameters, which may be used for discrete or continuous tracking of error syndromes in future implementation of the pair cat code.  

This Article is organized as the following: In Section II we 
discuss a model of the pair photon driven dissipation process in our circuit QED system while introducing our experimental setup.  In Section III we present experimental characterization of our PCS stabilization process.  This includes measurements of the pair photon population dynamics, direct PND measurements, a demonstration of manifold stabilization, and join Wigner tomography.  In Section IV we introduce and implement the subspace tomography that leads to quantitative understanding of the non-ideality of the stabilized PCS.  We conclude in Section V with a brief summary and outlook.

\section{PCS stabilization scheme in circuit QED}
Our approach to generate and stabilize a pair coherent state 
is based on the application of a pair photon drive counter-balanced  
by engineered pair photon dissipation 
and a cross Kerr interaction.  To understand how the PCS naturally emerges as the steady state under their combined effect, the two-mode system dynamics can be compared or mapped to: 1) the textbook example of the stabilized Glauber coherent state of a driven damped oscillator, and 2) previous demonstrations of stabilized single-mode cat-state manifolds~\cite{leghtas_confining_2015, grimm_stabilization_2020}.  The correspondence of relevant Hamiltonian and dissipator terms are listed in Table~\ref{table:PC_principle}.  

While the pair-photon drive $\hat{H}_d=\epsilon_{ab}\hat{a}^\dagger\hat{b}^\dagger+c.c.$, also known as the two-mode squeezing drive~\cite{heidmann_observation_1987}, has long been a workhorse in quantum optics, stabilization of the PCS requires either a strong (non-unitary) pair photon loss mechanism or a strong (unitary) cross-Kerr interaction (strong relative to the single-photon decay rates).  These two possible strategies are analogous to the stabilized ``dissipative cat"~\cite{leghtas_confining_2015,lescanne_exponential_2020} and ``Kerr cat"~\cite{puri_engineering_2017,grimm_stabilization_2020}, respectively, in the single-mode two-photon processes.  In fact, the dissipative and unitary effects play the roles of real and imaginary components of the restoring force and are mutually compatible (see Table~\ref{table:PC_principle}).  Inspired by both types of cat-state stabilization, our system combines both effects while generalizing this coherent stabilization to the two mode scenario. 
The 3d circuit QED architecture is ideal for realizing this driven dissipation process due to the availability of strong coupling between modes, the four-wave-mixing capability of the Josephson junction, and the wide range of mode lifetimes achievable in the same system.


\begin{figure}[tbp]
    \centering
    \includegraphics[scale=1]{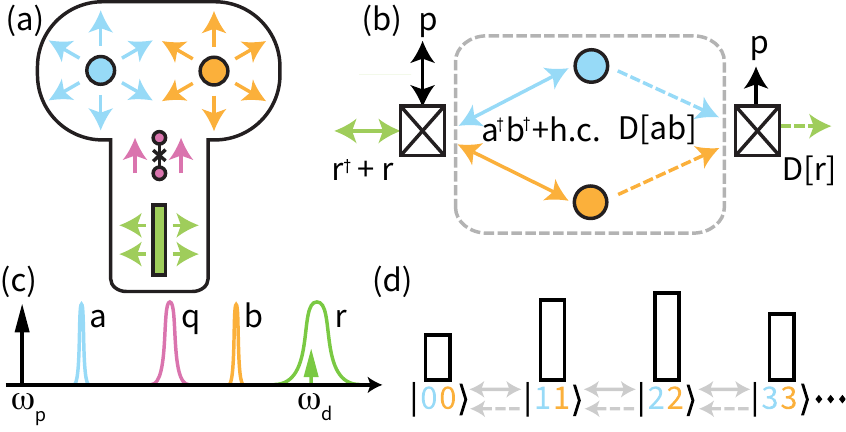}
    \caption{Pair coherent state generation: system and protocol. (a) Cartoon of the 3D cQED system containing two high-Q cylindrical post cavities $a$ and $b$ (blue and orange), the transmon ancilla $q$ (magenta), and the stripline low-Q reservoir $r$ (green). (b) Mixing drive $p$, with frequency $\omega_p \approx \omega_a + \omega_b - \omega_r$, coherently converts reservoir excitations with pairs of $a$ and $b$ excitations mediated by the ancilla junction. 
    The same ancilla junction also converts the reservoir decay (dotted green arrow) into an effective pair photon dissipative process (dotted blue/orange arrows).  After adiabatic elimination of the reservoir, the storage cavity dynamics are shown in the dotted box. (c) Cartoon of the mode frequencies and linewidths (not to scale). Strong off-resonance CW mixing drive $p$ and weak on-resonance ($\omega_d\approx\omega_r$) reservoir drive $d$ are shown as vertical arrows. (d) Schematic diagram of $\delta=0$ states under pair photon drive (double-headed arrow) and pair dissipation (dotted arrow). Fock states are written as $\ket{n_a n_b}$ and vertical bars represent a PCS distribution. 
}
\label{fig:PC_1}
\end{figure}

As shown in Fig.~\ref{fig:PC_1}(a), our system contains two cylindrical post cavity modes $a$ and $b$ (with single photon loss rates $\kappa_a/2\pi=0.30$ kHz and $\kappa_b/2\pi=0.74$ kHz), a stripline resonator $r$ (with decay rate $\kappa_{r}/2\pi = 0.78$ MHz) used for readout and as a Markovian reservoir, and an ancilla transmon $q$.  
The device architecture is similar to that in Ref.~\cite{wang_schrodinger_2016} except that the two cavity posts share the same elliptical cavity body to allow strong transmon-cavity couplings with a relatively small transmon antenna (see Appendix \ref{Setup}). The cavity/resonator modes have annihilation operators $\hat{a}$, $\hat{b}$, $\hat{r}$.  
The leading order terms of the static system Hamiltonian in the rotating frame are:
\begin{align}
    \hat{H_0} = \hat{H}_{\textrm{disp}}+\hat{H}_{\textrm{sk}}+\hat{H}_\textrm{rk}
    \label{eq:H0}
\end{align}
where 
\begin{align}
    \hat{H}_\textrm{disp}/\hbar =
    - \sum_{m=a,b,r}
     \big(\chi_{m}\ket{e}\bra{e}+\chi^{f}_{m}\ket{f}\bra{f}\big) \hat{m}^\dagger\hat{m}
\end{align} 
are the dispersive interaction terms between the lowest three transmon levels ($\ket{g}, \ket{e}, \ket{f}$) and the cavity/resonator modes, which we use for characterization of the cavity bosonic states, 
\begin{equation}
    \hat{H}_\textrm{sk}/\hbar = - K_{ab} \hat{a}^\dagger\hat{a} \hat{b}^\dagger\hat{b} - \frac{K_{aa}}{2} (\hat{a}^\dagger\hat{a})^2 - \frac{K_{bb}}{2} (\hat{b}^\dagger\hat{b})^2 
    \label{eq:H_SK}
\end{equation}
are the cross-Kerr and self-Kerr nonlinearities of the storage cavities which contributes to PCS stabilization, and
\begin{equation}
    \hat{H}_\textrm{rk}/\hbar = - (K_{ar} \hat{a}^\dagger\hat{a} + K_{br} \hat{b}^\dagger\hat{b}) \hat{r}^\dagger\hat{r} - \frac{K_{rr}}{2} (\hat{r}^\dagger\hat{r})^2
    \label{eq:H_RK}
\end{equation}
are the Kerr terms involving the reservoir mode, which are spurious nonlinearities in this experiment.  The device is measured in a dilution refrigerator at a nominal base temperature of 20 mK.  All device parameters are listed in the Appendix Table II, with the general rates hierarchy of $\chi_{m} \gg \kappa_r \gg K_{mn} \gg \kappa_{a,b}$ (where $m,n=a,b,r$).

\begin{figure*}[tbp]
    \centering
    \includegraphics[scale=0.8]{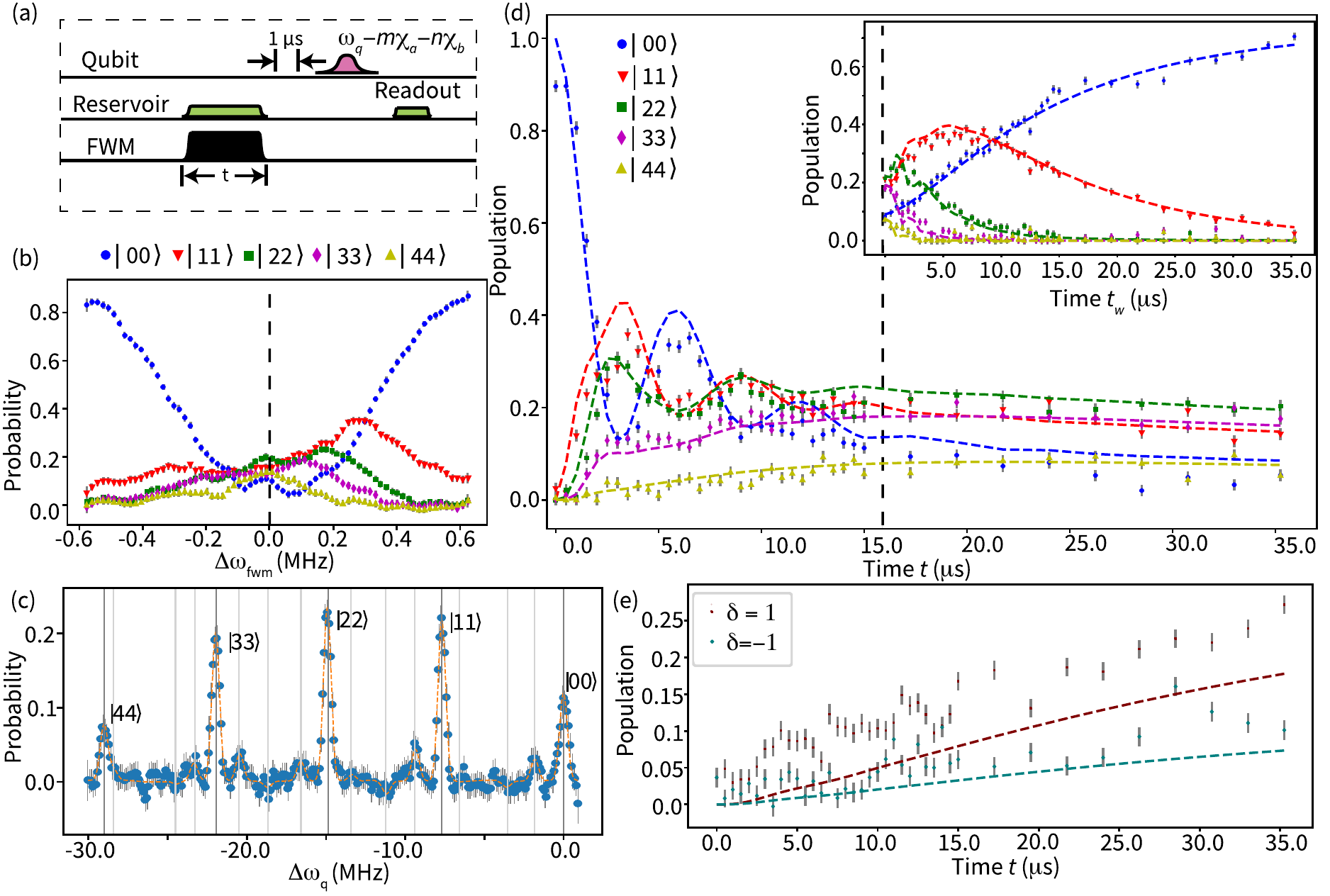}
    \caption{Pair coherent state characterization: population and time dynamics. (a) Pulse sequence using selective ancilla rotation to analyze photon population dynamics. We use an ancilla $\pi$-pulse with a frequency as shown to measure the population of a two-cavity state $\ket{mn}$ after waiting for 1 $\mu$s for the reservoir to relax. (b) Populations of two-cavity states $\ket{00}$ through $\ket{44}$ after $t=15$ $\mu$s of pumping at a different mixing pump frequency (while keeping the reservoir drive constant). The rest of the experiments are carried out under the condition shown as $\Delta\omega_\mathrm{fwm}=0$ here.  
    (c) Transmon spectroscopy after 15 $\mu$s of pumping with cw mixing and reservoir tones. Dark vertical bars correspond to $\delta=0$ states while light bars correspond to error states caused by single photon loss. (d) Time-domain photon population measurements performed starting in vacuum and pumping with both cw tones for a variable time $t$. 
    The inset shows a time-domain measurement of the pair photon decay (without pair photon drives) by keeping only the mixing pump on for a variable amount of time after the two-tone pumping for $t=15$ $\mu$s (vertical dashed line).  
    Dashed curves are numerical fits to the pair photon dynamics model Eq.~(\ref{eq:rho_full}) plus single-photon losses, including only two free parameters: $\epsilon_{ab}/2\pi = 99$ kHz and $\kappa_{ab} = (12.7$ $\mu s)^{-1}$. 
    (e) Measurement of the photon number difference $\delta$, where the probability of $\delta=\pm1$ is measured over variable time $t$. The dashed lines are simulation results with the same parameters as in (d). 
}
\label{fig:PC_2}
\end{figure*}

To implement the pair photon excitation and dissipation, we apply two stabilization drives to our system: a strong off-resonance pump 
that coherently converts reservoir photons with pairs of photons in $a$ and $b$, and a weaker drive approximately on resonance with the reservoir (Fig.~\ref{fig:PC_1}b,c). 
Under these two drives, the Hamiltonian gains an interaction term under the rotating wave approximation:
\begin{equation}
    \hat{H}_{int}/ \hbar = g_{ab} \hat{a}^\dagger \hat{b}^\dagger \hat{r} + \epsilon_d \hat{r}^\dagger + h.c.
    \label{eq:H_int}
\end{equation}
where $\epsilon_d$ is the rate of the reservoir drive, and $g_{ab}$ is the four-wave mixing rate activated by the off-resonance pump. If we assume the transmon remains in the ground state and then adiabatically eliminate the reservoir due to its fast relative dynamics, we obtain our desired form of the Lindblad master equation for the reduced density matrix $\rho$ of the two storage cavities: 
\begin{align}
     \frac{\partial \rho}{\partial t} = 
    -&\frac{i}{\hbar} \big[ (\epsilon_{ab} \hat{a}^\dagger\hat{b}^\dagger 
    + K_\text{eff} \hat{a}^\dagger\hat{a} \hat{b}^\dagger\hat{b} + h.c.), \rho\big]\nonumber\\
    &+\mathcal{D} \big[\sqrt{\kappa_{ab}} \hat{a}\hat{b}+\zeta_a\hat{a}^\dagger\hat{a}+\zeta_b\hat{b}^\dagger\hat{b}\big](\rho)
    \label{eq:rho_full}
\end{align}
where we have combined the storage Kerr terms ($\hat{H}_\textrm{sk}$) as a single $K_\text{eff}$ term (valid in the limit of $\gamma \gg \delta$).  $\mathcal{D}[\hat{o}](\rho)$ is the Lindblad superoperator with a composite jump operator $\hat{o}=\sqrt{\kappa_{ab}} \hat{a}\hat{b}+\zeta_a\hat{a}^\dagger\hat{a}+\zeta_b\hat{b}^\dagger\hat{b}$, where $\zeta_a$ and $\zeta_b$ are complex coefficients related to $\hat{H}_\mathrm{rk}$.  In the limit of $g_{ab}\gg K_{ar},K_{br}$, we have $\zeta_a, \zeta_b=0$, and this jump operator is reduced to the desirable form of two-photon loss, $\hat{o}\propto\hat{a}\hat{b}$, and the system is stabilized to the PCS manifold as described in Table~\ref{table:PC_principle}. 

Unlike previous analyses of two-photon driven dissipation~\cite{leghtas_confining_2015, lescanne_exponential_2020}, we emphasize that Eq.~(\ref{eq:rho_full}) is valid even when the reservoir mode is driven far away from vacuum (\textit{i.e.}~displaced to a large coherent state when $\epsilon_d > \kappa_r$) as long as the adiabatic condition, $\kappa_r\gg g_{ab}, K_{mn}$, is satisfied.  
The effective pair photon drive $\epsilon_{ab}$ and pair photon loss rate $\kappa_{ab}$ both increase with the four-wave mixing rate, $\epsilon_{ab} = -2ig_{ab} \epsilon_d/\kappa_r$, $\kappa_{ab} = 4 |g_{ab}|^2/\kappa_r$, while only the pair photon drive increases with the reservoir drive.  Initially unexpectedly, the reservoir nonlinearity, $\hat{H}_\textrm{rk}$, enters as dephasing-like modifications added to the pair-photon loss operator $\sqrt{\kappa_{ab}}\hat{a}\hat{b}$ in Eq.~(\ref{eq:rho_full}), which will contribute significantly to the experimental outcome.  See Appendix \ref{Derivation} for analytical derivation of the system dynamics. 

\section{General characterization of pair photon stabilization}
\subsection{Pair Photon Population Dynamics}
To characterize the pair photon driven dissipative dynamics, we first measure the two-cavity photon number distributions for any prepared state. To do this we perform spectroscopy of the ancilla transmon whose frequency is shifted by $-\chi_{a,b}$ for every photon in cavity $a$, $b$.  A frequency-selective rotation of the ancilla at a detuning $\Delta\omega_q = -n_a\chi_a-n_b\chi_b$ maps the probability of being in Fock state $\ket{n_a n_b}$ to the ancilla excitation, which can then be read out.  Here, we have written the two-cavity Fock state $\ket{n}_a\otimes\ket{m}_b$ as $\ket{nm}$ in short, a convention that will be used for the rest of the Article. To find an optimal condition to create a PCS in the presence of ac-Stark shift, we sweep the frequency of the four-wave mixing pump while measuring populations of various $\delta=0$ states (Fig.~\ref{fig:PC_2}b).
After 15 $\mu$s of pumping with calibrated drive rates of $g_{ab}/2\pi=60$ kHz and $\epsilon_{d}/2\pi=780$ kHz, we observe the full spectroscopy of the ancilla (Fig.~\ref{fig:PC_2}c) which illustrates the Poisson-like distribution of a PCS.  We can also track these photon populations over time to understand how the system converges to a quasi-steady state with a pair photon drive and pair photon dissipation (Fig.~\ref{fig:PC_2}d) or how it decays under the pair photon dissipation alone (inset of Fig.~\ref{fig:PC_2}d). 

Fitting these time-domain data of pair photon population dynamics to numerical simulations of Eq.~(\ref{eq:rho_full}) (Appendix \ref{Simulations}), we extract the pair photon dissipation rate $\kappa_{ab}/2\pi=12.5$ kHz (from Fig.~\ref{fig:PC_2}d inset) and subsequently the pair photon drive rate $\epsilon_{ab}/2\pi=99$ kHz (from Fig.~\ref{fig:PC_2}d), in good agreement with values expected from adiabatic elimination.  Together with the effective cross-Kerr of the system ($K_\textrm{eff}/2\pi=-86$ kHz), these rates are significantly faster than the undesirable single photon loss rates $\kappa_a$ and $\kappa_b$.  Therefore, we experimentally create a quasi-steady state resembling a PCS with $|\gamma|=2.3$ and $\delta=0$ under these driven dissipation conditions before single photon loss eventually decoheres the state by altering $\delta$. 

We note that different combinations of $n_a$ and $n_b$ can in principle result in similar dispersive shift to the ancilla, causing ambiguity in our measurement of cavity photon population.  However, for our system with $3\chi_{qa} \approx \chi_{qb}$, this ambiguity arises only when the underlying Fock state $\ket{n_a n_b}$ deviates from its expected $\delta$ by at least 4, which is highly unlikely to occur over time scales shorter than the single photon losses. 

A prominent feature of the population dynamics in Fig.~\ref{fig:PC_2}(d) is oscillations between states which are then damped to a steady state.   The oscillations arise from the under-damped Kerr dynamics of the storage cavities, $\kappa_{ab}<|K_\textrm{eff}|$, and are due to the sudden turn-on of the stabilization drives not allowing the storage cavities to adiabatically evolve in the ground state of the Kerr Hamiltonian as in Ref.~\cite{grimm_stabilization_2020}. Nevertheless, the pair photon dissipation plays a crucial role in relaxing the system to the steady state, allowing us to by-pass the otherwise slow ramping of the pump tones required for adiabatic state preparation.  This hybrid implementation of dissipation and Hamiltonian stabilization is analogous to recent proposals of combining Kerr cat qubits with engineered dissipation to further improve robustness against unwanted excitations~\cite{gautier_combined_2022, guillaud_repetition_2019, putterman_stabilizing_2022}.  The spurious dephasing-like contributions in the jump operator in Eq.~(\ref{eq:rho_full}) also contributes to the convergence to a steady state, which, however, is accompanied with loss of coherence as will be discussed later.

\subsection{Measurement of Photon Number Difference}
Single photon loss is a fundamental decoherence channel in superconducting cavities which is not stabilized by the pair-photon driven dissipation.   Quantum non-demolition measurement of photon number difference (PND) followed by autonomous or digital feedback is a crucial component of the pair-cat QEC code~\cite{albert_pair-cat_2019}. 
An existing proposal of PND measurement requires an exact negative $\chi$-matching condition, $\chi_{a} = -\chi_{b}$, 
which allows direct mapping of the probability of the two-cavity state in a targeted $\delta$ onto the ancilla excitations via a simple selective rotation~\cite{albert_pair-cat_2019}.  
For a transmon ancilla, this negative $\chi$ matching condition does not occur naturally but can be achieved with additional strong off-resonant pumps~\cite{rosenblum_cnot_2018}.  It should be noted that managing multiple strong off-resonant pumps in cQED devices without incurring instabilities remains an active challenge~\cite{lescanne_escape_2019}.

Here we demonstrate an alternative method of PND measurement without $\chi$ matching: We apply a comb of number-selective $\pi$-pulses to excite the ancilla at frequencies corresponding to the dispersive shifts for all cavity states (below a reasonable truncation) with a targeted $\delta$. 
Subsequent readout of the ancilla informs whether the PND of the two-cavity state is equal to $\delta$ or not.
For instance, we could superimpose pulses at $\omega_q-n\chi_{a}-(n+1)\chi_{b}$, where $0\leq n \leq 5$, to inquire whether the system is in $\delta=1$. In Fig.~\ref{fig:PC_2}(c) we use this measurement to probe the populations in the $\delta=\pm 1$ states, which shows the effect of single photon loss.  While our experimental setup does not have a quantum-limited amplifier to perform the necessary demonstration, this technique of PND measurement is in principle quantum non-demolition as long as the phases of the $\pi$-pulses are tuned to be effectively equal after compensating for any ac Stark shift and Kerr effects.  It can then be used to repetitively monitor single photon loss from a state with known $\delta$ 
so long as $\ket{00}$, $\ket{10}$, and $\ket{01}$ are number resolved and a recovery operation can be applied before the next photon loss occurs.  Therefore, it provides a practical avenue for QEC of single photon loss in the pair cat code, hence relaxing one of its demanding requirements for implementation.  The comb-based PND measurement may also be converted to an autonomous QEC protocol to correct photon loss errors~\cite{gertler_protecting_2021}. 

\begin{figure}[t]
    \centering
    \includegraphics[scale=1.01]{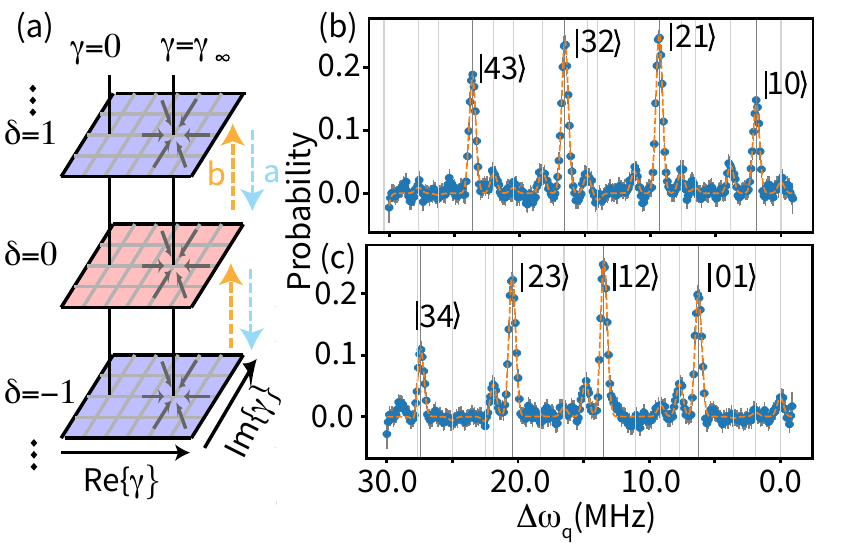}
    \caption{The manifold of stabilized pair coherent states. (a) Conceptual representation of the different $\delta$ subspaces of the two-cavity system under pair photon driven dissipation.  Each subspace has a unique steady state $\ket{\gamma_\infty, \delta}$, which functions as an attractor in a generalized phase plane. In general, the system is stabilized to the manifold of states represented by the vertical string $\gamma=\gamma_\infty$.  
    (b, c) Ancilla spectroscopy that demonstrates the photon number distribution of the stabilized (b) $\delta=1$ and (c) $\delta=-1$ PCS.  These measurements are performed by preparing initial states of $\ket{10}$ and $\ket{01}$ respectively and then applying the exact same stabilization drives as used in Fig.~\ref{fig:PC_2}(c) for $t=15$ $\mu$s.   
    For each plot, dark vertical lines indicate the desired states while light vertical lines mark error states due to single photon loss or imperfect initial state preparation. }
    \label{fig:PC_3}
\end{figure}

\subsection{Manifold Stabilization}

An important property of the pair photon driven dissipation process is conservation of the photon number difference $\delta$: 
it stabilizes $\gamma$ while allowing $\delta$ to be the one and only degree of freedom inherited from arbitrary initial values and may allow quantum operations in the presence of stabilizing pumps. For example, the photon number distribution in Fig.~\ref{fig:PC_2} corresponds to the unique steady state of the two-cavity system within the $\delta=0$ subspace, and its $\delta$ is inherited from the initial vacuum state before the pair-photon pumping is applied.  To create a PCS with $\delta=1$ or $-1$, we can use a Selective Number-dependent Arbitrary Phase (SNAP) gate~\cite{heeres_cavity_2015} to prepare a $\ket{10}$ or $\ket{01}$ initial state before applying the same pumping conditions.  The resultant photon number distributions are shown in Fig.~\ref{fig:PC_3}, corresponding to PCS of $\ket{\delta =1, |\gamma|\approx2.4}$ and $\ket{\delta =-1, |\gamma|\approx2.2}$ respectively.
More generally, the pair photon dynamics of Eq.~(\ref{eq:rho_full}), in the ideal limit of $\zeta_a,\zeta_b=0$, confines the two-cavity state to a quantum manifold (Hilbert subspace) spanned by a series of PCS with a fixed $\gamma$ and an arbitrary integer $\delta$.  Any coherent superpositions of these PCS are steady states allowed by the driven dissipation although the coherence is protected against single photon loss, which shift the value of $\delta$ between neighboring integers.  

\begin{figure*}[tbp]
    \centering
    \includegraphics[scale=.78]{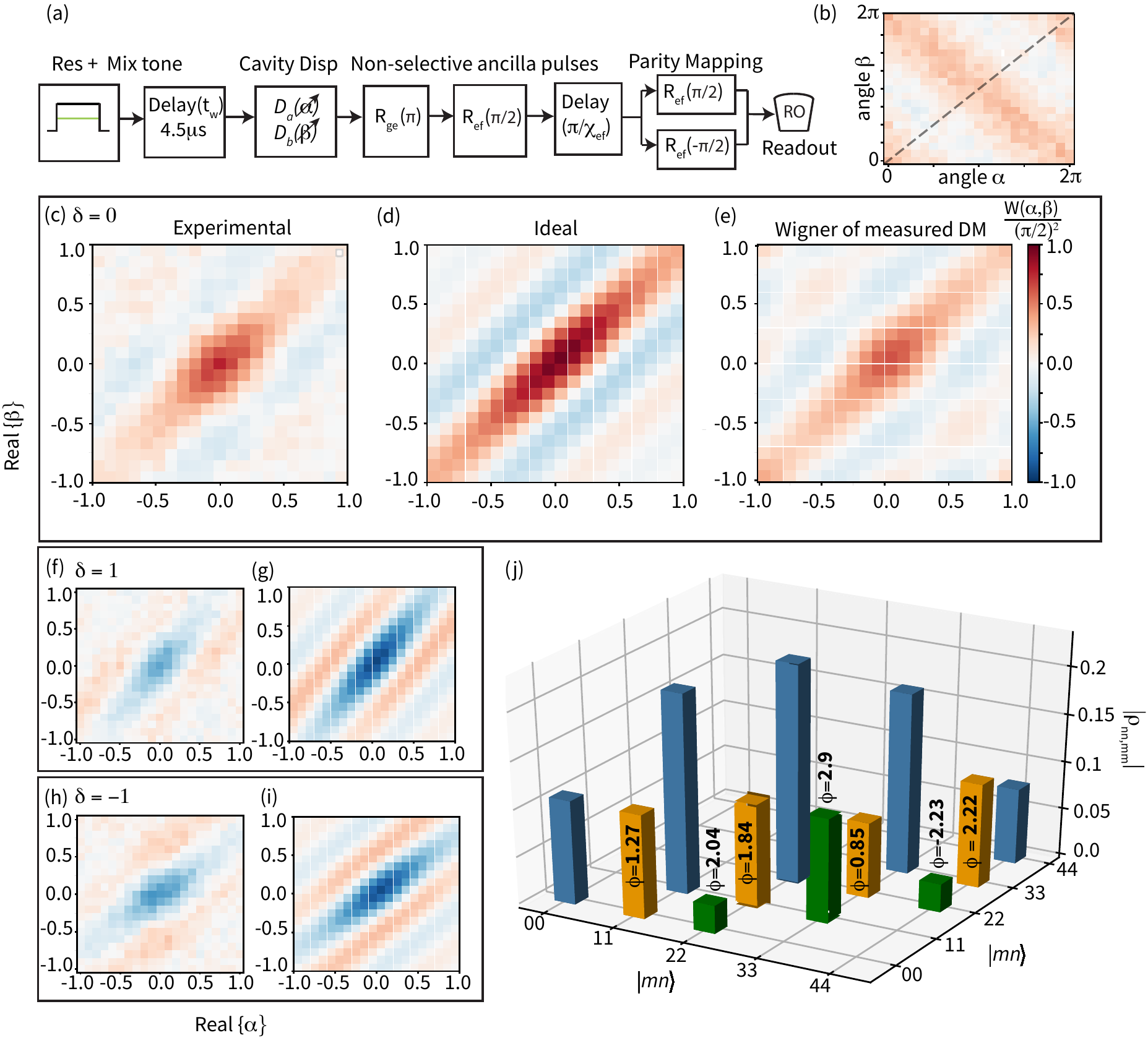}
    
    \caption{Joint Wigner tomography of pair coherent states. (a) The pulse sequence for Wigner function measurements. The two cw tones are applied for $t=15$ $\mu$s to prepare and stabilize a state that is then evolved under no pump tones for $t_w=4.5$ $\mu$s to reach the first phase revival of an approximate PCS under cavity Kerr dynamics (and for the reservoir to relax to vacuum).  The joint parity mapping uses a Ramsey sequence with evolution time of $\pi/\chi_{ef}$, where $\chi_{ef}\approx\chi_a^f-\chi_a\approx\chi_b^f-\chi_b$.  (b) Wigner function, $\frac{\pi^2}{4}W(\alpha, \beta)$, of the stabilized $\delta=0$ state with constant displacement amplitudes $|\alpha| = |\beta| = 0.3$ and sweeping over the angle of both displacements. 
    (c) Measured Re$[\alpha]$-Re$[\beta]$ cut of the joint Wigner function for the approximate $\ket{\gamma=2.3, \delta=0}$ PCS (with the real axes redefined such that $\arg[\gamma]=0$).  This can be compared to the same Wigner function cuts of (d) the ideal PCS $\ket{2.3, 0}$ and (e) the experimentally measured density matrix block [as shown in (j)] of this stabilized state. 
    (f), (h) Measured Re$[\alpha]$-Re$[\beta]$ cut of the joint Wigner function for the approximate $\ket{\gamma=2.4, \delta=1}$ and $\ket{\gamma=2.2, \delta=-1}$ PCS stabilized under the same condition, to be compared to (g), (i) the corresponding ideal PCS respectively. (j) The $\delta=0$ sector of the density matrix of the stabilized $\delta=0$ state at the same effective free evolution time, measured with quantum subspace tomography (\textit{Note:} independent from the Wigner measurements) as discussed in Section IV. The blue bars are the diagonal elements indicating population distribution whereas the off-diagonal coherence elements are shown by orange (green) for nearest (next-nearest) neighboring Fock-state pairs. Other coherence elements are small and hence not measured in the experiment.  Phases of the coherence elements are shown numerically over the bars. 
}\label{fig:PC_4}
\end{figure*}

\subsection{Joint Wigner Measurements}
To demonstrate the non-classicality and coherence of these stabilized states, we measure the joint Wigner function of the two-cavity state $\rho$~\cite{cahill_density_1969}:
\begin{equation}
    W(\alpha,\beta) = \frac{4}{\pi^2}\textrm{Tr}[\rho \hat{D}_a(\alpha)\hat{D}_b(\beta)\hat{P}_J\hat{D}^\dagger_b(\beta)\hat{D}^\dagger_a(\alpha)]
\end{equation}
where $\hat{P}_{J} = e^{\pi i (\hat{a}^\dagger\hat{a}+\hat{b}^\dagger\hat{b})}$ is the joint photon number parity operator, and $\hat{D}_a(\alpha),\hat{D}_b(\beta)$ are phase-space displacement operators of the two cavities respectively.  We take advantage of our approximately matched dispersive shift of the $\ket{e}-\ket{f}$ transmon transition: $\chi^f_a-\chi_a=\chi^f_b-\chi_b$ to measure the joint photon number parity, whose expectation value following cavity displacements in the 4D phase space ($\alpha,\beta$) can be directly scaled to the joint Wigner function~\cite{wang_schrodinger_2016, milman_proposal_2005}.  

In Fig.~\ref{fig:PC_4}(b), with a 2d angular cut of the joint Wigner function at fixed displacement amplitudes, we first demonstrate an interesting property of the PCS (or any eigenstates of PND in general): invariance with respect to the differential cavity phase: $W(\alpha,\beta)=W(\alpha e^{i\phi},\beta e^{-i\phi})$.  This is because the product of cavity rotation operators $\hat{R}_a(\phi)\hat{R}_b(-\phi)=e^{i\phi\hat{a}^\dagger\hat{a}}e^{-i\phi\hat{b}^\dagger\hat{b}}=e^{i\phi\hat{\delta}}\equiv\mathbb{I}$ for a PCS up to a global phase.  On the other hand, oscillation with respect to the total phase is a signature of phase coherence of the two-cavity state.  Choosing the common phase $\frac{1}{2}(\arg[\alpha]+\arg[\beta])$ where the Wigner function takes the maximum value, we show measured 2d phase-space cuts of the joint Wigner function of the experimental $\delta=0$, +1, and -1 PCS to be compared to the closest-matched ideal PCS (Fig.~\ref{fig:PC_4}(d-i)).  These Wigner data are acquired after $t=15$ $\mu$s of stabilizing drives and an additional $t_w=4.5$ $\mu$s of wait time.  We note that $t_w\gtrsim 5/\kappa_r=1$ $\mu$s is necessary to allow the reservoir photons to fully decay, and the wait time here is chosen to reach the first phase revival of an approximate PCS after cavity Kerr dynamics~\cite{kirchmair_observation_2013}.  The characteristic interference fringes of the PCS are clearly visible, indicating appreciable coherence and consistent multi-photon phases.  The even-parity ($\delta=0$) and odd-parity ($\delta=\pm1$) states show opposite Wigner contrasts as expected, and the differences in the slope of the fringes are striking features indicating the different PND of the states.

The technique of direct joint Wigner tomography in principle allows full reconstruction of the two-cavity density matrix but faces a multitude of practical challenges.  
Firstly, tomographic reconstruction requires a very large amount of data.  Even if we impose a low cutoff of 4 photons per cavity, the tomography still requires accurate measurements at a minimum of 625 (typically many more) different phase-space sampling points. 
Secondly, the joint parity measurement contains some photon-number-dependent systematic errors caused by the storage-readout cross-Kerr, the 6th-order dispersive shift of the transmon, and non-perfect $\chi$-matching for joint parity extraction.  Last but not least, even with sufficient brute force, it is challenging to prevent small stochastic and systematic errors from propagating badly in the matrix pseudo-inversion problem of reconstruction due to the presence of singularities.  In our experiment, primarily to limit spurious readout signals due to a large $K_{br}$, we have limited our cavity displacements to $|\alpha|, |\beta|<1$ in Wigner function measurements as in Fig.~4, which contains nearly all the salient features of the PCS.  However, a full Wigner reconstruction would require cumulative measurements of highly diluted features over a much larger extent of the 4D phase space.  

\begin{figure*}[tbp]
    \centering
    \includegraphics[scale=.35]{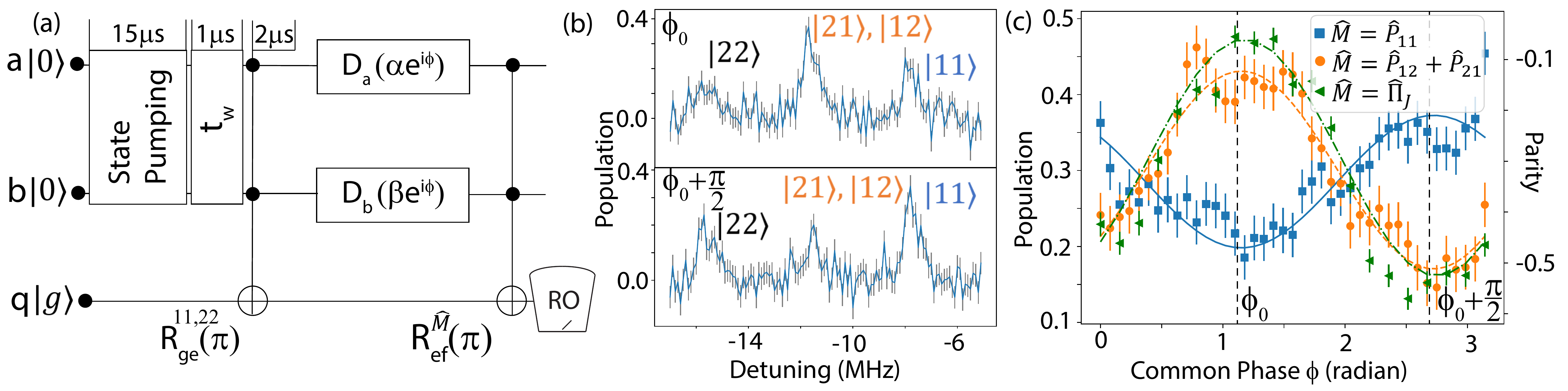}
    \caption{Subspace tomography protocol. (a) Example schematic for tomography in subspace $S=\textrm{span}\{\ket{1,1},\ket{2,2}\}$.  Number-selective transmon $\pi$-pulses (2 $\mu$s-long Gaussian) are applied  to isolate the two Fock components in the $\ket{e}$ manifold.  Cavity displacement pulses are then applied to induce interference, followed by mapping an observable $\hat{M}$ (e.g. the joint parity) to $\ket{f}$.   
    (b) Spectroscopy of the $\ket{e}-\ket{f}$ transition of the transmon after the cavity displacements $\hat{D}_a(|\alpha|e^{i\phi}) \hat{D}_b(|\beta| e^{i\phi})$ 
    in (a) with two different displacement phases $\phi = \phi_0$ and $\phi_0+\pi/2$ where $\phi_0$ is chosen to minimize the $\hat{M}=\hat{P}_{11}$ measurement, thus $\phi_0+\pi/2$ will maximize it. The difference in photon number distribution population in the two cases demonstrates the presence of interference.  The $\ket{21}$ and $\ket{12}$ peaks overlap significantly due to the approximately matched dispersive shift of the $\ket{e}-\ket{f}$ transmon transition.   
    (c) The population of $\ket{11}$, the sum of the populations of $\ket{21}$ and $\ket{12}$, and the photon number parity of the displaced state in the $\ket{e}$ manifold, measured as a function of the common displacement phase $\phi$.  All fit curves correspond to the scaled coherence element $\rho_{11,22}/\sqrt{\rho_{11}\rho_{22}} = 0.63 e^{0.84i}$. 
    } 
    \label{fig:subspace_tomo}
\end{figure*}

\section{Quantum Subspace tomography}

Quantitative insight into our experimental two-cavity state is enabled by a new characterization tool focusing on specific subspaces of interest.  The goal is to completely characterize the projected density matrix $\rho_{SS}=\hat{P}\rho \hat{P}$, with projection operator $\hat{P}$ for subspace $S$ of the system.  For $d$-dimensional subspace $S$, we only need to perform $d^2$ measurements for subspace tomography.  Using a three-level ancilla, one can effectively isolate $S$ from the rest of the Hilbert space using the $\ket{g}-\ket{e}$ transition of the ancilla, and using the $\ket{e}-\ket{f}$ transition to perform tomography.  A general protocol for this method is discussed in Appendix \ref{Protocol}.  

\subsection{Implementation of 2d Subspace Tomography}
We demonstrate this subspace tomography technique with a direct and self-calibrated measurement of the pair-wise coherence between the constituent states of the stabilized $\delta=0$ state.  This is realized by entangling only two Fock components, e.g.~$\ket{11}$ and $\ket{22}$, with $\ket{e}$ of the transmon using number-selective $\pi$-pulses.  Within this 2d subspace, since the diagonal elements of the density matrix, which we denote as $\rho_{11}$ and $\rho_{22}$, have been measured via transmon spectroscopy (Fig.~2), the complex-valued off-diagonal element $\rho_{11,22}$ is the only unknown.  
We then displace both cavities $\hat{D}_a(|\alpha|e^{i\phi}) \hat{D}_b(|\beta| e^{i\phi})$ with fixed amplitudes (typically $|\alpha|$, $|\beta|<0.5$) and vary the common phase $\phi$, 
which redistributes the photon populations in the $\ket{e}$ manifold and shows constructive or destructive interference in each Fock component depending on $\rho_{11,22}$ and $\phi$ [Fig.~\ref{fig:subspace_tomo}(b)].  
Using number-selective $\ket{e}$-$\ket{f}$ rotation of the transmon, we can measure the population of any individual target state after displacement, such as $\ket{11e}$, which oscillates as a function of $\phi$. Alternatively, we can also measure the oscillation in other observables, such as the joint parity [Fig.~\ref{fig:subspace_tomo}(c)].  The amplitude and phase of such oscillations can be directly converted to $\rho_{11,22}$ following a comparison with easily calculable properties of the state $D(\alpha,\beta)(\sqrt{\rho_{11}}\ket{11}+\sqrt{\rho_{22}}\ket{22})$. 
Repeating the same procedure for different pairs of states, we obtain the $\delta=0$ block of the system density matrix by direct measurements of its individual off-diagonal elements [Fig.~4(f)].  In practice, we run numerical calculation to find the optimal $|\alpha|$, $|\beta|$ and measurement observable for each pair of states to maximize the visibility of the oscillations. 

\begin{figure*}[tbp]
    \centering
    \includegraphics[scale=.75]{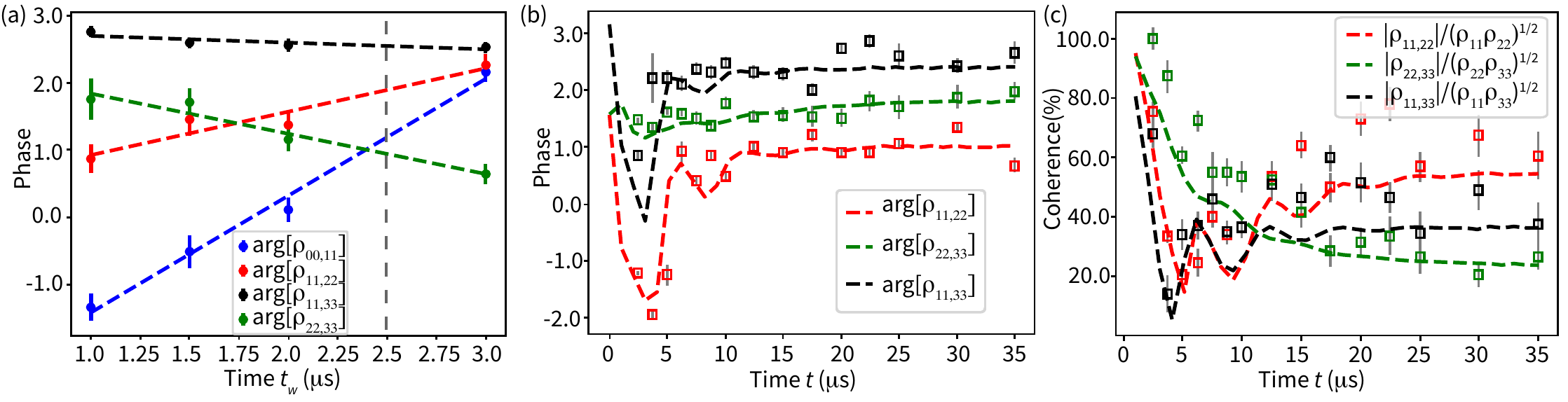}
    \caption{
    (a) Phases of selected off-diagonal density matrix elements of the two-cavity state, measured with the $d=2$ subspace tomography, as a function of wait time $t_w$ after the state has been prepared by the stabilizing drives for $t=15$ $\mu$s.  The phase evolution agrees with the cavity Kerr terms $\hat{H}_{sk}$ and the cavity detuning in the drive frame. Dashed lines are linear fits, which allow us to interpolate or extrapolate the phases of these states to different $t_w$. The vertical dashed line corresponds to the time at which the density matrix sector is plotted in Fig.~\ref{fig:PC_4}(j).  In order to compare with Wigner data in Fig.~\ref{fig:PC_4}, note that the effective free-evolution time of the two-cavity state here is $t_w+2$ $\mu$s, accounting for the duration of the selective transmon $\ket{g}-\ket{e}$ pulses in the subspace tomography protocol (Fig.~\ref{fig:subspace_tomo}(a)). 
    (b, c) Phases and relative coherence of selected off-diagonal density matrix elements of the two-cavity state for different stabilization pumping time $t$, measured with subspace tomography after $t_w=1$ $\mu$s. 
    Dashed lines: numerical simulation results of the storage pair photon dynamics according to Eq.~(\ref{eq:rho_full}), which does not contain the reservoir mode and captures all prominent features of the data.   
    \label{fig:coh_over_time}
}
\end{figure*}

Experimental implementation of the subspace tomography technique requires careful tracking of rotating frames and phase accumulation of different branches of states.  To ensure self-consistent phase measurements, we work in the two-cavity rotating frame set by two cw stabilization drives (the drive frame), and require the four frequencies of the tomographic pulses precisely sum up to 0 (Appendix Fig.~\ref{fig:phase_loop}).  Since the pulse frequencies for selective transmon rotations 
have no freedom for adjustment, we apply our fast cavity displacement drives at deliberately detuned frequencies in order to exactly compensate for the irregular detunings of all other tones due to various Stark shifts and higher-order nonlinearity of the system (Appendix \ref{Freq-matching}).  This strategy allows us to measure the phases of off-diagonal density matrix elements that are unambiguously defined relative to the two stabilization drives.  As shown in Fig.~\ref{fig:coh_over_time}(a), following $t=15$ $\mu$s pumping, our measured phases of $\rho_{00,11}, \rho_{11,22}, \rho_{22,33}, \rho_{33,44}$ all evolve over wait time $t_w$ as expected from the frequencies of the corresponding transitions in the drive frame.  Note that their frequencies differ due to the Kerr Hamiltonian of the storage cavities [Eq.~(\ref{eq:H_SK})].   Fig.~\ref{fig:coh_over_time}(b) further shows that the measured phases are constant over extended pumping times $t$. 

A key advantage of this subspace tomography technique of $d=2$ is that it maps individual density matrix elements of interest to individual experimental signatures in an intuitive manner, making the experimental uncertainties of the tomographic measurements highly transparent.  For example, from the sinusoidal fits in Fig.~\ref{fig:subspace_tomo}(b), we can unambiguously report the uncertainly of the amplitude (3\%) and phase (0.05 radian) for $\rho_{12,21}$, 
which would be extremely challenging to obtain in traditional tomographic reconstruction. 
Furthermore, it is relatively straightforward to extend our scheme to $d>2$ by applying more than 2 selective ancilla $\pi$-pulses in isolating the subspace, although extra care make be taken in accounting for phase accumulations if the selective pulses are not exactly equally spaced in frequency.  

\subsection{Understanding Deviations from Ideal PCS}

With the capability of direct pair-wise coherence measurements, we can track the amplitude and phase of selected coherence elements in the two-cavity state as a function of stabilization pumping time $t$.  We find that the cavities indeed converge to a quasi-steady state (except for the slow process of single photon loss) with persistent coherence (Fig.~\ref{fig:coh_over_time}(c)). However, the magnitude of the steady-state coherence is far lower than the ideal values $|\rho_{nn,mm}|=\sqrt{\rho_{nn}\rho_{mm}}$.  Moreover, the stabilized cavity phases, which we 
can extrapolate backwards in time from data such as Fig.~\ref{fig:coh_over_time}(a), also differ considerably from the ideal PCS, which should have equal superposition phases between neighboring Fock components, i.e.~$\arg[\rho_{nn,(n+1)(n+1)}]=\arg[\gamma]$).  Due to the Kerr evolution of the storage cavities during idle time ($K_\textrm{eff}=-86$ kHz), we expect a stabilized PCS to collapse and revive every 5.8 $\mu$s~\cite{kirchmair_observation_2013}, making $t_w=5.8$ $\mu$s the first optimal timing to observe the PCS in Wigner measurements.  However, the measured pair-wise phases of the stabilized state suggests that the two-cavity state most resembles a PCS at $t_w\approx4.5$ $\mu$s, which has been confirmed by the joint Wigner measurements.  This skewness in phases and loss of coherence is caused by the relatively large cross-Kerr term of the reservoir mode, $K_{br}/2\pi = 58$ kHz, which is comparable to the competing mixing drive rate $g_{ab}/2\pi$ (while  $K_{ar}/2\pi = 7$ kHz is much smaller).

Remarkably, this spurious effect from the reservoir cross-Kerr can be well captured by the simple reduced model of the pair-photon dynamics Eq.~(\ref{eq:rho_full}), which features a composite jump operator $\hat{o}=\sqrt{\kappa_{ab}} \hat{a}\hat{b}+\zeta_a\hat{a}^\dagger\hat{a}+\zeta_b\hat{b}^\dagger\hat{b}$.  Through adiabatic elimination of the $r$ mode (Appendix B), we find $\zeta_b=\sqrt{4K^2_{br}r_0^2/\kappa_r}$ (and similarly for $\zeta_a$), where $r_0$ is the complex displacement amplitude of the $r$ mode during the stabilization process.  This coefficient is identical to that of photon shot-noise dephasing of qubits~\cite{gambetta_qubit-photon_2006}, with $K_{br}$ taking the place of the qubit-cavity dispersive shift.  Indeed, if the mixing drive is turned off (hence $\kappa_{ab}=0$), any initial storage cavity state $\rho$ should be dephased by the fluctuating photon numbers in a driven $r$ mode, which can be described by the usual dephasing jump operators $\hat{a}^\dagger\hat{a}$ and $\hat{b}^\dagger\hat{b}$.  In our experiment, $|r_0|=1.8$, 
which gives rise to a dephasing rate $|\zeta_b|^2/2\approx 2\pi\cdot28$ kHz while the effect of $\zeta_a$ is much less significant. Intuitively, the finite steady-state coherence of the system in Fig.~\ref{fig:coh_over_time} can be understood from the competition between the PCS stabilization mechanism and the spurious dephasing effects.

It is important to note that the pair-photon dissipation and the two dephasing-like terms constitute a single jump operator $\hat{o}$, not separate ones.  This is because the three events are all directly originating from the same stochastic event of reservoir photon loss.  As a result, the relative phases between the terms in $\hat{o}$ play a crucial role in skewing the stabilized state from the ideal PCS, including a larger state size, a broader distribution of photon numbers, and the varying phases between neighboring Fock components.  
A numerical simulation of Eq.~(\ref{eq:rho_full}) using this composite jump operator reproduces the prominent coherence elements of the pumped two-cavity state, which is in good agreement with the experimental data in Fig.~\ref{fig:coh_over_time} with no free parameters except for a two-cavity global frame rotation of 1.5 radian (which can be attributed to the difference in cable electrical length between the stabilization drive lines and the cavity displacement lines together with a small additional cavity rotation during the reservoir ring-down before the tomography).

To improve the quality of the stabilized PCS, it is necessary to suppress the spurious dephasing terms in the composite jump operator $\hat{o}$. Ideally, this can be accomplished by engineering a Josephson circuit without the reservoir cross-Kerr, analogous to the asymmetrically threaded SQUID circuit for dissipative cat-state stabilization~\cite{lescanne_exponential_2020}.  In our current transmon-based device, this may be accomplished by either an extra Kerr-cancellation pump tone~\cite{zhang_drive-induced_2022} or by detuning the two stabilization drives a few (reservoir) linewidths away to reduce photon number fluctuations in the $r$ mode.  The former will also suppress $|K_\textrm{eff}|$ and hence boosts the state size and the dissipative nature of the stabilization scheme.  The latter will suppress $\kappa_{ab}$ and move our PCS stabilization deeper into the regime relying on Hamiltonian confinement.  Given the full understanding of the pair photon dynamics at the conclusion of this experiment, we believe both strategies can yield improved fidelity for the stabilized PCS in our device, although new forms of nonlinear ancillas are needed to realize spurious-free pair photon processes at faster rates.

\section{Summary and outlook}
There is a growing history of circuit QED experiments inspired by the field of quantum optics.  The convenience of microwave systems and the powerful nonlinearity of superconducting circuitry has paved new ways 
to the study of exotic bosonic states that may be otherwise prohibitively challenging to implement.  
Owing to the identical nature of bosons, bosonic quantum states offer the opportunities of hardware-efficient dissipation engineering schemes, but also poses unique challenges requiring new characterization techniques.  Both aspects are reflected in our experiments.  Unlike earlier demonstrations of two-mode non-Gaussian states such as the entangled cat state~\cite{wang_schrodinger_2016} or the N00N state~\cite{wang_deterministic_2011}, the PCS should contain more than one e-bit (EPR pair) of entanglement.  It would be interesting to further develop efficient tools to calibrate and distill the entanglement in such states. 

The pair cat code, a recent addition to the zoo of bosonic QEC codes, offers the tantalizing prospect to correct both photon loss and dephasing errors (to the first order) autonomously and fault-tolerantly~\cite{albert_pair-cat_2019}.  
Our experimental realization of the PCS, demonstration of the first cross cavity dissipator, and introduction of a convenient (although not fault-tolerant) PND measurement are all valuable steps towards a pair cat code.  
However, fully implementing and utilizing the advantages of the pair cat code will require schemes to suppress forward propagation of ancilla errors and upgrading the pair photon dissipation to a four photon cross cavity dissipator, which involves a leap in experimental complexity.  

As a first attempt towards a two-mode bosonic code, our work suggests several challenges moving from one mode to two.  Unlike the concatenation of a smaller multi-qubit QEC code to a larger one, two-mode bosonic code states generally cannot be created from concatenating single-cavity bosonic qubits using cavity-cavity logic gates.  Generation of two-mode states in circuit QED requires specific cross-cavity nonlinear interactions, which has been implemented in our work with a relatively simple hardware setup but not without spurious effects and difficult parameter trade-offs.  Future experiments will have to tailor the cross-cavity nonlinear interactions likely using Josephson circuits with a strong 3rd-order nonlinearity~\cite{bergeal_phase-preserving_2010} while suppressing unwanted 4th-order terms.  
In addition, since two-mode states reside in a much larger Hilbert space, their implementation requires both effective confinement to preferred subspaces and efficient methods to diagnose imperfections.  To the latter, our work demonstrates a valuable tool in subspace tomography, which provides a general framework for future experiments to isolate key components of a high-dimensional quantum state for characterization.

\subsection*{Acknowledgements}

We thank Michel Devoret, Ioannis Tsioutsios, Akshay Koottandavida and Yu-Xin Wang for helpful discussions.  We thank Juliang Li and Xiaowei Deng for experimental assistance.  This research was supported by the US Department of Energy (DE-SC0021099).  Fabrication and initial characterization of the device was supported by the Air Force Office of Scientific Research (FA9550-18-1-0092). LJ would like to acknowledge the support from the ARO (W911NF-18-1-0212), 
AFOSR MURI (FA9550-19-1-0399) and the Packard Foundation (2020-71479).\\

\noindent
$\dagger$ There authors contributed equally.

\noindent
$*$ Correspondence and requests for materials should be addressed to Chen Wang, wangc@umass.edu.

\begin{appendix}

\renewcommand\thefigure{\arabic{figure}}  
\renewcommand{\figurename}{\textbf{Appendix Fig.}}
\renewcommand{\tablename}{\textbf{Appendix Table}}

\section{Experimental Setup and Methods}\label{Setup}
\subsection{Device architecture and transmon fabrication}
The system consists of two post cavities dispersively coupled to a fixed frequency transmon ancilla which is then dispersively coupled to a stripline $\lambda/2$ resonator that is used for readout and as reservoir in the experiment. The posts are made with high-quality Aluminum inside a cavity as shown in Fig. ~\ref{fig:fridge_alCav}b. The cavity has a tunnel opening on one of the sides which is where the sapphire chip containing the qubit and resonator is held with a chip clamp. This high-quality superconducting cavity is made with 5N ($99.999\%$) purity Aluminum and it shields the qubit from magnetic field lines that can degrade the coherence. The ancilla qubit and the cavity modes each have their own drive ports. The readout signal is collected from another port that is coupled strongly with the stripline resonator. The coupling strength between all the modes and the coherence numbers for the modes are listed as in the Table. ~\ref{table:parameters}. The setup with the cavities and chip was initially simulated in the electromagnetic solver software called Ansys HFSS to design the frequencies of the modes and interactions between them that are needed for the experiment. 

\par The simulated chip design is fabricated in a clean room facility for nanofabrication using the standard procedure. We use a 30keV JEOL JSM-7001F SEM to perform electron beam lithography to define the transmon and the stripline resonator in one step.  The aluminum thin film is evaporated on the sapphire chip using Plassys MEB550S.  The transmon has a single Al-AlO$_x$-Al Josephson junction produced by the Dolan bridge method where the aluminum is deposited in two different angles while allowing some time in between for it to oxidize.  

\subsection{Measurement setup}
\begin{figure*}[tbp]
    \centering
    \includegraphics[scale=0.6]{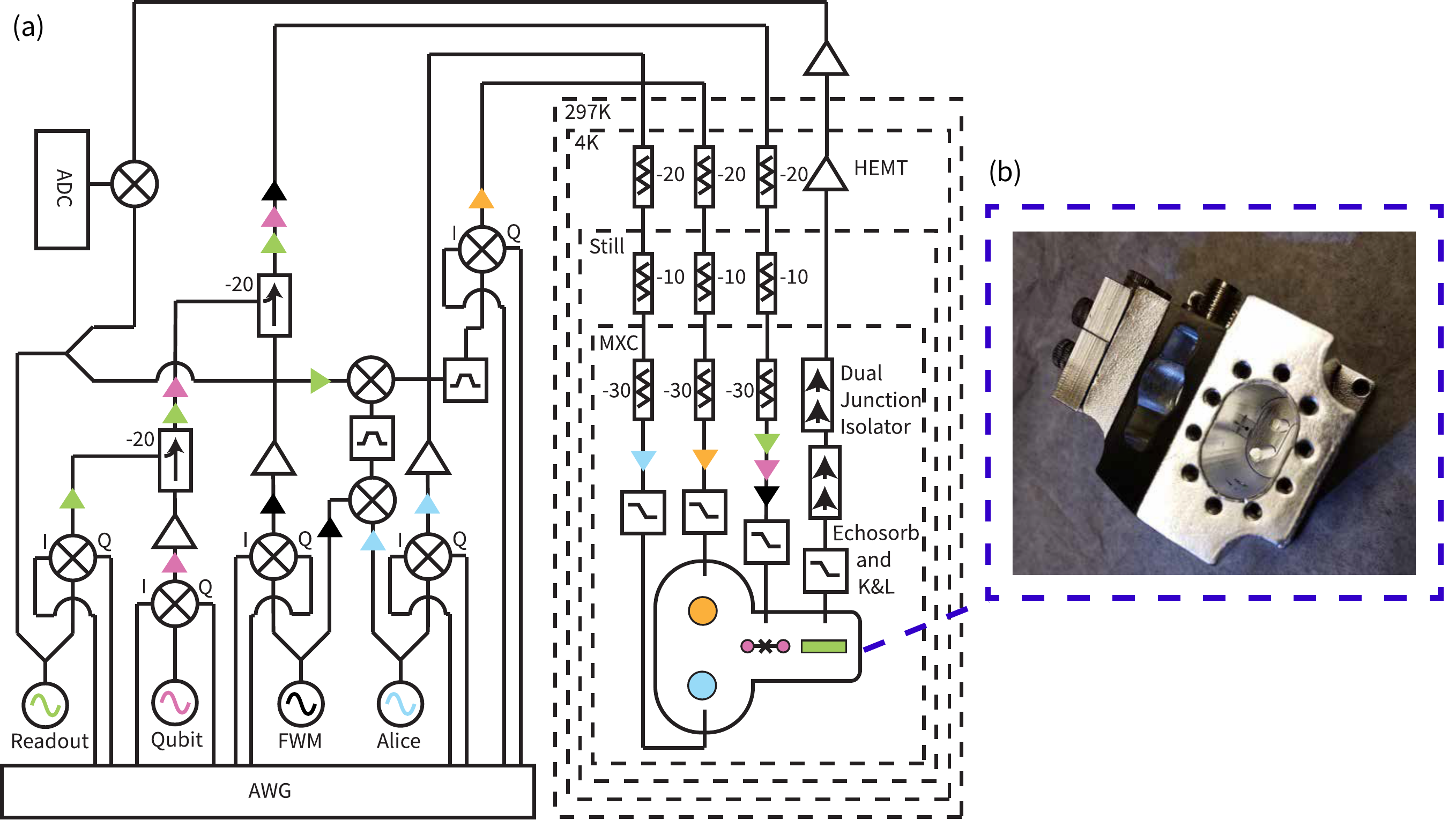}
    \caption{(a) Fridge and Measurement Wiring Diagram: We have four signal generators which, with eight AWG channels, make input signals for all the modes and FWM tone. As discussed in the supplementary, the measurement setup shows that the drive signal for Cavity $b$ is generated by mixing the signals from cavity $a$, reservoir and the FWM generators with 3-port mixers and using the correct sideband filters. (b) Picture of high purity aluminum cavity used in the experiment. The posts that give the storage cavity modes are seen inside the aluminum body with the sapphire chip entering the cavity from a side tunnel. The chip is held by an aluminum clamp and the transmon qubit antenna is at equal distance from both the posts. 
    The cavity is mounted on a bracket that is placed inside a high-permeability magnetic shield thermalized to the mixing chamber.     
}\label{fig:fridge_alCav}
\end{figure*}

The measurement setup is designed to generate microwave signals that drive all the four modes in the system and to receive signals from the readout port that are digitized to read the qubit state. We use the basic microwave engineering technique of mixing signals from the signal generator and the arbitrary waveform generator (AWG) using an IQ-mixer to modulate the sidebands. The filtered RF output from each mixer then goes into the fridge to the input ports as shown in Fig. ~\ref{fig:fridge_alCav}a. The signal collected from the readout port goes to a 3-port mixer for demodulation and then to a digitizer. In addition to drives for the modes of the system the setup also produces continuous four-wave mixing tone and a reservoir tone that satisfy the frequency matching condition as discussed in the main text to create a PCS. Inside the fridge all the input signals are attenuated by 20 dBm at 4K plate, 10 dBm at Still plate and 30 dBm at the MXC, which then go through the eccosorbs and last level of filtering at MXC before going into the input ports. 
\par We want to study the coherence of the prepared entangled state in the storage cavities by probing them to measure 2D slices of joint wigners as shown in Fig.~\ref{fig:PC_4}b-d. Using different generators to pump and probe the cavities would cause the phase of the state in the storage cavities to not be locked to the cavity displacement drives used to measure Wigner functions, thus smearing out any phase coherence. We therefore eliminate a phase degree of freedom by generating cavity B drive from a reservoir, cavity A and the four-wave mixing generators. As shown in Fig.~\ref{fig:fridge_alCav}a the signal from the generators for the four-wave mixing pump and cavity A drive are mixed with a 3-port mixer which, with a low pass filter, produces an output signal at $\omega_{a}-\omega_{p}$. This signal is again mixed with the signal from the reservoir drive generator with a 3-port mixer and a low pass filter on the output to produce $\omega_{r}-\omega_{a}+\omega_{p}$, thus generating a local oscillator for the IQ-mixer modulating the sideband for driving cavity B.      

\begin{table}[bp]
\caption{Measured System parameters.\\ 
$^\dagger$ Values measured in the presence of the relatively strong four-wave mixing tone. \\
$^*$ $K_{aa}$ is estimated from the measured $K_{ab}$ and $K_{bb}$. 
}
\centering
\begin{tabular}{c c c c c}
\hline\hline\\[-2ex]
		& Symbol	&  Value	 \\
\hline\\[-2ex]
Transmon frequency  & $\omega_q/2\pi$	& 5378 MHz\\
Transmon anharmonicity  &  $\alpha_q/2\pi$	& 204 MHz \\
Transmon $T_1$  & $T_{1q}$	& $40$ $\mu$s \\
Transmon $T_2^*$ Ramsey  & $T^*_{2q}$ & $18$ $\mu$s \\
Transmon $T_2$ Echo  & $T_{2q}$	& $60$ $\mu$s \\
Transmon $\ket{e}_q$ population  & & 2\% $(3.5\%^{\dagger})$ \\
\hline\\[-2ex]
Reservoir frequency  & $\omega_r/2\pi$ & 7409 MHz \\

Reservoir-transmon coupling  & $\chi_{qr}/2\pi$ & 3.09 MHz \\

Reservoir anharmonicity & $K_{rr}/2\pi$ & 12 kHz \\
Reservoir decay rate  & $\kappa/2\pi$ & 0.78 MHz \\
\hline\\[-2ex]
Cavity $a$ frequency & $\omega_a/2\pi$ & 4072 MHz \\
Cavity $a$-transmon coupling & $\chi_{qa}/2\pi$ & 1.89 MHz \\
Cavity $a$ anharmonicity & $K_{aa}/2\pi$ & 8 kHz$^{*}$ \\
Cavity $a$ $T_1$ & $T_{1a}$ & $530$ $\mu$s \\
Cavity $a$ $T_2$ & $T_{2a}$ & $400$ $\mu$s \\
Cavity $a$ $\ket{1}_a$ population &  & $\sim$2\%\\
\hline\\[-2ex]
Cavity $b$ frequency & $\omega_b/2\pi$ & 6094 MHz \\
Cavity $b$-transmon coupling & $\chi_q/2\pi$ & 6.26 MHz \\
Cavity $b$ anharmonicity & $K_{bb}/2\pi$ & 71 kHz (81 kHz$^{\dagger}$)\\
Cavity $b$ $T_1$ & $T_{1b}$ & $216$ $\mu$s \\
Cavity $b$ $T_2$ & $T_{2b}$ & $200$ $\mu$s \\
Cavity $b$ $\ket{1}_b$ population &  & $\sim$1\%\\
\hline\\[-2ex]
Cavity $a$-$b$ cross-Kerr & $K_{ab}/2\pi$ & 48 kHz (53 kHz$^{\dagger}$)\\
Cavity $a$-$r$ cross-Kerr & $K_{ar}/2\pi$ & 7 kHz\\
Cavity $b$-$r$ cross-Kerr & $K_{br}/2\pi$ & 58 kHz\\
\hline
\end{tabular}
\label{table:parameters}
\end{table}

\subsection{Pump Tuneup Procedure}

Given the presence of the storage-reservoir cross-Kerr interaction, there is no perfect experimental pumping condition that can exactly stabilize the PCS.  In our experimental procedure, with the intention of studying driven dissipative dynamics, we aimed to maximize the pair-photon dissipation rate $\kappa_{ab}$ while empirically obtaining a steady state with relatively large photon numbers and minimal undesirable heating effects.  

In order to maximize the ratio of the two-photon dissipation rate to the single-photon loss rate in the cavities $\kappa_{ab}/\kappa_{a,b}$, a relatively strong four-wave mixing (FWM) pump amplitude $g_{ab}$ is needed.  We set the power of the FWM pump at a value where we start to observe a small measurable rise of the excited state population of the ancilla transmon.  The power of the reservoir drive is set to displace the reservoir to a few photons in its steady state, as can be measured via the Stark shift and dephasing of the transmon. We start by setting the frequency of the reservoir and FWM drive to be on resonance with the reservoir $\omega_{r}$ and the FWM condition $\omega_{p}=\omega_{a}+\omega_{b}-\omega_{r}$, respectively.  We also use spectroscopy measurements to calibrate the Stark-shifted mode frequencies, which give us improved estimates of the pumping frequency condition.  
 These two crudely set tones allow preparation of some non-trivial 
 pair photon states (e.g.~with substantial $\ket{11}$ and $\ket{22}$ states) after 10-20 $\mu$s of pumping time.  

To calibrate the optimal frequency for the FWM pump that maximizes $\kappa_{ab}$, we start with the above crudely-prepared state, and then attempt to apply only the mixing pump at varying frequencies to evacuate photons in pairs.  We finely sweep through the FWM frequencies, and measure the population of $\ket{00}$ after a fixed amount of pumping time.  The FWM detuning $\Delta_{p}$ at which we get the largest $\ket{00}$ state gives the maximum $\kappa_{ab}$.  Experimentally we found $\Delta_{p}=-125$ kHz relative to expected value from undriven mode frequencies $\omega_{p}=\omega_{a}+\omega_{b}-\omega_{r}$, primarily due to the Stark shift.  
We can extract the two-photon dissipation rate $\kappa_{ab}$ from these decay measurements by fitting them to simulations as discussed in Appendix C.

The next step is to tune the amplitude and frequency of the reservoir drive to make a relatively large $\delta=0$ pair photon state with Poisson-like distribution of Fock state amplitudes in line with the PCS distribution.  
We found that the optimal reservoir drive frequency is about -185 kHz detuned from its bare frequency, which can be explained by the cross-Kerr effect $\hat{H}_{rk}$ in the presence of storage photons and the ac Stark from the pump tone.  
Finally, in a verification experiment depicted in Fig.~\ref{fig:PC_2}(b), we measure the cavity photon population distribution after 15 $\mu$s of stabilization drives with a fixed reservoir drive condition while varying FWM frequency.  This figure shows that indeed the pair-photon drive is also most efficient when we pump at the FWM frequency tuned up to the previously determined $\Delta_{p}$ based on maximizing pair-photon dissipation.

\section{Driven Hamiltonian and Adiabatic elimination}\label{Derivation}

\subsection{Rotating frame Hamiltonian transformation}

In this section, we derive the four-wave mixing effect of the off-resonant pump tone on the Josephson circuit under the rotating wave approximation, arriving at the drive Hamiltonian in the rotating frame Eq.~(\ref{eq:H_int}).  This procedure is similar to various previous experiments~\cite{leghtas_confining_2015, gao_programmable_2018, gertler_protecting_2021}.  

We start by writing our Hamiltonian as a sum of our four modes (reservoir: $r$, qubit: $q$, storage a: $a$, storage b: $b$) coupled to a josephson junction with two drives (FWM pump: $p$, reservoir drive: $d$) applied to the reservoir mode~\cite{blais_circuit_2020}:
\begin{align}
     \hat{H}_{full}/\hbar = \sum_{m=q,r,a,b}\bar{\omega_{m}}\hat{m}^\dagger\hat{m} - 
     E_{J}(cos(\hat{\phi})+\hat{\phi}^{2}/2) \\\nonumber +2Re(\epsilon_{p}e^{-i\omega_{p}t}+\epsilon_{d}e^{-i\omega_{d}t})(\hat{r}+\hat{r}^\dagger)
\end{align}
\begin{equation}
     \hat{\phi} = \sum_{m=q,r,a,b}\phi_{m}(\hat{m}^\dagger+\hat{m})
\end{equation}
The first term in Eq.~(B1) represents the linear character of each mode corresponding to $a_{i}$. The Josephson junction is represented by the cosine term with $E_{J}$ as the Josephson energy and $\hat{\phi}$ as the phase across the junction decomposed into the contributions from each mode with contribution of $\phi_{m}$ to the zero point fluctuations of $\hat{\phi}$. The two drive terms are acting on the reservoir mode with complex amplitudes $\epsilon_{d}$, $\epsilon_{p}$ and frequencies $\omega_{d}$, $\omega_{p}$, respectively, where the pump is a strong off resonant tone and the drive is a weak near resonant tone with the reservoir mode.

We are in a regime where the following inequality holds:
\begin{equation}
     \omega_{p},\omega_{d},\bar{\omega_{m}} \gg  \epsilon_{p} \gg \frac{E_{J}}{\hbar} ||\hat{\phi}||^{4}/4!
\end{equation}
We now want to make a change of frame using the following unitary to eliminate the fastest time scales:
\begin{equation}
     U = e^{i\bar{\omega_{q}}t\hat{q}^\dagger\hat{q}}e^{i\omega_{d}t\hat{r}^\dagger\hat{r}}e^{i\omega_{da}t\hat{a}^\dagger\hat{a}}e^{i\omega_{db}t\hat{b}^\dagger\hat{b}}e^{-\Tilde{\xi_p}\hat{r}^\dagger+\Tilde{\xi_p}^{*}\hat{r}}
\end{equation}
We are rotating out the qubit frequency $\bar{\omega_{q}}$ simply for convenience as it will not change the rest of the analysis in any meaningful way. We then rotate out the drive frequency $\omega_{d}$ on the reservoir mode so the drive term is stationary in the chosen frame. We then want to put our storage modes in the drive frame of $\omega_{d}+\omega_{p}$. We can do this by rotating out frequency $\omega_{da}(\omega_{db})$ from mode $a(b)$. Any $\omega_{da},\omega_{da}$ that satisfy $\omega_{
da}+\omega_{db} = \omega_{d}+\omega_{p}$ will put us in the correct drive frame, thus there is some freedom as to which frequencies we choose. Here we choose a constant offset($\frac{\Delta_{sd}}{2}$) of $\omega_{da},\omega_{db}$ from $\bar{\omega_{a}},\bar{\omega_{b}}$, respectively, such that $\bar{\omega_{a}} -\omega_{da} = \bar{\omega_{b}} - \omega_{db} = \frac{\Delta_{sd}}{2}$. Finally we want to apply a displacement unitary to eliminate the time dependant amplitude ($\Tilde{\xi_{p}}$) in the reservoir caused by the pump tone to bring the effect of this amplitude into the $\hat{\phi}$ operator allowing the effects on the other modes to become directly apparent. Looking at times on the order or greater than $1/\kappa_{r}$ we can ignore the reservoir transient dynamics and look at the steady state response of our displacement amplitude as done in \cite{leghtas_confining_2015}:
\begin{align}
     \Tilde{\xi_{p}}&=\xi_{p}e^{-i\omega_{p}t},\\\nonumber
     \xi_{p} &= \frac{-i\epsilon_{p}}{\kappa_{r}/2 + i(\bar{\omega_{r}}-\omega_{p})}\approx \frac{-i\epsilon_{p}}{\kappa_{r}/2 + i(\omega_{r}-\omega_{p})}
\end{align}
The Hamiltonian in this new frame ($\hat{H}_{full}'$) now becomes:
\begin{align}
     \hat{H}_{full}'/\hbar =& (\bar{\omega_{r}}-\omega_{d})\hat{r}^\dagger\hat{r}+\Delta_{sd}\hat{a}^\dagger\hat{a}\\\nonumber&+\Delta_{sd}\hat{b}^\dagger\hat{b} - \frac{E_{J}}{\hbar}(cos(\Tilde{\hat{\phi}})+\Tilde{\hat{\phi}}^{2}/2)
\end{align}
\begin{align}
     \Tilde{\hat{\phi}} = \sum_{m=q,r,a,b}\phi_{m}(\Tilde{\hat{m}}+\Tilde{\hat{m}}^\dagger) + (\Tilde{\xi_{p}}+\Tilde{\xi_{p}}^{*})\phi_r
\end{align}
\begin{align}
     &\Tilde{\hat{q}}=\hat{q}e^{-i\bar{\omega_{q}}t},\Tilde{\hat{r}}=\hat{r}e^{-i\omega_{d}t}, \\\nonumber &\Tilde{\hat{a}}=\hat{a}e^{-i\omega_{da}t},\Tilde{\hat{b}}=\hat{b}e^{-i\omega_{db}t}
\end{align}
We now expand the cosine term to fourth order and keep only non-rotating terms in alignment with the rotating wave approximation and separate the Hamiltonian into 3 parts ($\hat{H}_{full}' = \hat{H}_{freq} + \hat{H}_{Kerr} + \hat{H}_{drive}$) as defined below:
\begin{align}
    \hat{H}_{freq} = & (- \chi_{qr}|\xi_{p}|^{2})\hat{q}^\dagger\hat{q} \\\nonumber
    & + (\omega_{r}-\omega_{d}-2K_{rr}|\xi_{p}|^{2})\hat{r}^\dagger\hat{r}\\\nonumber
    & + (\Delta_{sd}-K_{ar}|\xi_{p}|^{2})\hat{a}^\dagger\hat{a}\\\nonumber
    & + (\Delta_{sd}-K_{br}|\xi_{p}|^{2})\hat{b}^\dagger\hat{b}
\end{align}
\begin{align}
    \hat{H}_{Kerr} =& \sum_{m=r,a,b}-\frac{K_{mm}}{2}{(\hat{m}^\dagger\hat{m})^{2}}-\alpha_{q}(\hat{q}^\dagger\hat{q})^{2}-\chi_{qr}\hat{q}^\dagger\hat{q}\hat{r}^\dagger\hat{r} \nonumber\\
    &-K_{ar}\hat{a}^\dagger\hat{a}\hat{r}^\dagger\hat{r}-K_{br}\hat{b}^\dagger\hat{b}\hat{r}^\dagger\hat{r}-K_{ab}\hat{b}^\dagger\hat{b}\hat{a}^\dagger\hat{a}
\end{align}
\begin{align}
    \hat{H}_{int} = g_{ab}\hat{a}^\dagger\hat{b}^\dagger\hat{r} + \epsilon_{d}\hat{r}^\dagger + h.c.
\end{align}
$H_{freq}$ gives the frequency shifts to all the elements and the shifts from terms containing $|\xi_{p}|^2$ are the AC Stark shift effects from the pump tone. $H_{Kerr}$ corresponds to both self and cross-Kerr coupling terms where $K_{mm}=\frac{E_{J}}{\hbar}\phi_{m}^{4}/2, K_{mm'}=\frac{E_{J}}{\hbar}\phi_{m}^{2}\phi_{m'}^{2}$ where $m\neq m'$ and $g_{ab} = \phi_a \phi_b \phi_r^{2} \xi_p$.Note $H_{0}$ in Eq.~(\ref{eq:H0}) is analogous to $H_{Kerr}$ just with a truncation of the allowed qubit states to the $\ket{f}$ state.

\subsection{Semi-classical analysis of reservoir state}

In this section we 
perform a semi-classical analysis of the Langevin equations of motion for the reservoir and storage modes and show that the impact of the storage states on the reservoir dynamics is quite small and warrants treating the effect as a small perturbation.  This condition forms the basis for 
the adiabatic elimination of the reservoir mode in the following section.

We first will assume the qubit is in the ground state during the stabilization. This leads us to work with the Hamiltonian:
\begin{align}
    \hat{H}=& \Delta_d\hat{r}^\dagger\hat{r} + \Delta_a\hat{a}^\dagger\hat{a}
    + \Delta_b\hat{b}^\dagger\hat{b} \\\nonumber &- \sum_{m=r,a,b}\frac{K_{mm}}{2}{(\hat{m}^\dagger\hat{m})^{2}} - K_{ab}\hat{a}^\dagger\hat{a}\hat{b}^\dagger\hat{b} \\\nonumber &- K_{ar}\hat{a}^\dagger\hat{a}\hat{r}^\dagger\hat{r} - K_{br}\hat{a}^\dagger\hat{a}\hat{r}^\dagger\hat{r} + \hat{H}_{int}
\end{align}
where $\Delta_d = (\omega_{r}-\omega_{d}-2K_{rr}|\xi_{p}|^{2}),\Delta_a = (\Delta_{sd}-K_{ar}|\xi_{p}|^{2}),\Delta_b = (\Delta_{sd}-K_{br}|\xi_{p}|^{2})$.  For the following analysis, for simplicity we will focus on analyzing the $\delta = 0$ storage state, but the conclusion, as will be explained later, is general.  In this state our $\hat{a}^\dagger\hat{a}$ and $\hat{b}^\dagger\hat{b}$ terms are equivalent, thus we can group the storage cavity self and cross Kerr terms into one Kerr term, $K_\textrm{eff}\hat{a}^\dagger\hat{a}\hat{b}^\dagger\hat{b}, K_\textrm{eff}=-K_{ab}-K_{aa}/2-K_{bb}/2$. This yields the following simplified Hamiltonian ($\hat{H}_{\delta 0}$) for a $\delta = 0$ state: 
\begin{align}
    \label{eq:Hd0}
    \hat{H}_{\delta 0}=& \Delta_d\hat{r}^\dagger\hat{r} + \Delta_a\hat{a}^\dagger\hat{a}
    + \Delta_b\hat{b}^\dagger\hat{b} \\\nonumber &- \frac{K_{rr}}{2}{(\hat{r}^\dagger\hat{r})^{2}} + K_\mathrm{eff}\hat{a}^\dagger\hat{a}\hat{b}^\dagger\hat{b} \\\nonumber &- K_{ar}\hat{a}^\dagger\hat{a}\hat{r}^\dagger\hat{r} - K_{br}\hat{a}^\dagger\hat{a}\hat{r}^\dagger\hat{r} + \hat{H}_{int}
\end{align}

We now want to motivate the statement that the effect of the storage modes on the reservoir is quite small from both the effects of the pumping $(g_{ab})$ and the reservoir storage cross Kerr $(K_{ar},K_{br})$. 
The pumping effects can be seen by looking at the Langevin equations for the classical amplitude analogs of $\langle\hat{r}\rangle \rightarrow r, \sqrt{\langle \hat{a}\hat{b}\rangle}\rightarrow s$ in this Hamiltonian. We can then get the steady state solutions to these Langevin equations:
\begin{align}
    \label{eq:Langevin}
    \frac{d}{dt}r & = -i\epsilon_d-(\kappa_r/2+i\Delta_{d})r-ig^*_{ab}s^2=0 \\\nonumber 
    \frac{d}{dt}s & = -2ig_{ab}s^*r+2iK_\mathrm{eff}|s|^2s=0
\end{align}
From this we can algebraically solve for the steady-state mode amplitude:
\begin{align}
    r = \frac{\epsilon_d}{-\frac{|g_{ab}|^2}{K_\mathrm{eff}}+\frac{i(\kappa_r+2i\Delta_d)}{2}}
\end{align}
We are in the limit of $|\kappa_r+2i\Delta_d| \gg \frac{2|g_{ab}|^2}{K_\mathrm{eff}}$ so we can look at the mode amplitude as an uncoupled driven mode with a perturbation from the coupling to the storage cavity as:
\begin{align}
    \label{eq:r0_def}
    & r \approx r_0 + \delta_{disp}\\\nonumber
    & r_0 = \frac{2\epsilon_d}{i(\kappa_r+2i\Delta_d)} \\\nonumber
    & \delta_{disp}  = -\frac{4\epsilon_d|g_{ab}|^2}{K_\textrm{eff}(\kappa_r+2i\Delta_d)^2}, |\delta_{disp}| \ll 1  
\end{align}
For our experimental parameters, $|r_0|=1.8$.  We note that the storage-reservoir Langevin equation Eq.~(\ref{eq:Langevin}) takes a similar form as previous single-mode two-photon driven dissipation experiments~\cite{leghtas_confining_2015, lescanne_exponential_2020}, but our system is in a different parameter regime due to the presence of a large $K_\mathrm{eff}$.  In the absence of $K_\mathrm{eff}$, the steady-state $r$ would be close to 0 even if $|r_0|>1$.

While this was derived under the steady state and with a $\delta=0$ state to more concretely illustrate the point, one can more generally consider the `pull' the storage modes have on the reservoir as $|g_{ab}|^2/K_\mathrm{eff}$ and compare this to the loss rate $\kappa_{r}$ and see that we are in a regime where $\kappa_{r}\gg|g_{ab}|^2/K_\mathrm{eff}$, thus the effect of the storage modes on the reservoir state via the pumping effects is quite small. Now the reservoir storage cross Kerr effects must be taken into account. The cross Kerr values will simply add a further detuning on top of $\Delta_{d}$ to the reservoir amplitude for each number state of the storage.  Since our pair coherent state amplitude $\gamma$ is on the order of the photon number in each mode in the system for the $\delta = 0$ state we are analyzing, we will use this to represent the average photon number in the inequality. Thus, we must satisfy the inequality $\kappa_{r} \gg \gamma(K_{ar}+K_{br})$ to ensure the effects of the storage reservoir cross kerr can be treated as a perturbation on top of the bare reservoir amplitude.  Since our parameter regime satisfies $\kappa_{r} \gg \gamma(K_{ar}+K_{br}),|g_{ab}|^2/K_\mathrm{eff}$, we are well motivated to treat the reservoir amplitude as an uncoupled amplitude $r_{0}$ with a perturbation added to account for the effect from the storage modes.

\subsection{Adiabatic elimination of the reservoir mode}
In this section we use the above analysis of the reservoir dynamics as an uncoupled amplitude with a small perturbation accounting for the storage mode effects to adiabatically eliminate the reservoir mode. We will do this by going into a displaced frame of the reservoir mode using the uncoupled mode ampliude $r_{0}$ in which the resulting reservoir state is solely from the perturbation effects of the storage mode and hence remains close to vacuum. This small resulting reservoir amplitude allows us to write the density matrix as a perturbative expansion in the reservoir states. Then using the fast time scale of $\kappa_{r}$ we can adiabatically eliminate the reservoir mode to arrive at Eq.~(\ref{eq:rho_full}).

Using the above analysis of the reservoir mode, we can apply a unitary transformation $U_{disp}=e^{-r_{0}\hat{r}^\dagger+r_{0}^{*}\hat{r}}$ to $\hat{H}$ resulting in $\hat{H}' = U_{disp}\hat{H}U_{disp}^\dagger$ to go into a frame where the coherent state displacement on the reservoir mode is represented by just $\delta_{disp}$ in addition to the effect on the amplitude from the reservoir storage cross kerr effects which are also small:
\begin{align}
    \hat{H}' =&
    \Delta_d(\hat{r}^\dagger\hat{r} + \hat{r}^\dagger r_{0} +\hat{r}r_{0}^* +|r_0|^2) + \Delta_a\hat{a}^\dagger\hat{a}
    \\\nonumber&+ \Delta_b\hat{b}^\dagger\hat{b} - \sum_{m=r,a,b}\frac{K_{mm}}{2}{(\hat{m}^\dagger\hat{m})^{2}} - K_{ab}\hat{a}^\dagger\hat{a}\hat{b}^\dagger\hat{b}\\\nonumber & -(K_{ar}\hat{a}^\dagger\hat{a} + K_{br}\hat{b}^\dagger\hat{b})(\hat{r}^\dagger\hat{r} + \hat{r}^\dagger r_{0} +\hat{r}r_{0}^* +|r_0|^2) \\\nonumber &+ g_{ab}\hat{a}^\dagger\hat{b}^\dagger(\hat{r}+r_{0}) + \epsilon_{d}(\hat{r}^\dagger+r_{0}^*) \\\nonumber &+ g_{ab}^*\hat{a}\hat{b}(\hat{r}^\dagger+r_{0}^*) + \epsilon_{d}^*(\hat{r}+
    r_{0})
\end{align}
One can first omit all constant energy offset terms and then combine the $\Delta_{d}r_{0}\hat{r}^\dagger + h.c.$ terms with the resulting $\kappa_r$ Lindbladian loss operator terms of the same order from the same unitary transformation to cancel out the $\epsilon_{d}$ terms, yielding the following simplified $\hat{H}'$:
\begin{align}
    \hat{H}' =&
    \Delta_d\hat{r}^\dagger\hat{r} + \Delta_a\hat{a}^\dagger\hat{a}
    + \Delta_b\hat{b}^\dagger\hat{b} \\\nonumber& - \sum_{m=r,a,b}\frac{K_{mm}}{2}{(\hat{m}^\dagger\hat{m})^{2}} - K_{ab}\hat{a}^\dagger\hat{a}\hat{b}^\dagger\hat{b}\\\nonumber& - (K_{ar}\hat{a}^\dagger\hat{a} + K_{br}\hat{b}^\dagger\hat{b})(\hat{r}^\dagger\hat{r} + \hat{r}^\dagger r_{0} +\hat{r}r_{0}^* +|r_0|^2) \\\nonumber& + g_{ab}\hat{a}^\dagger\hat{b}^\dagger(\hat{r}+r_{0})  + g_{ab}^*\hat{a}\hat{b}(\hat{r}^\dagger+r_{0}^*)
\end{align}
Now, we can group terms that are small and say that the following dimensionless quantities,$\frac{\gamma(K_{ar}+K_{br})}{\kappa_{r}},\delta_{disp},\kappa_{a(b)}/\kappa_{r},g_{ab}/\kappa_{r},K_{aa(bb)}/\kappa_{r}$, are on the order of $\epsilon$ where $\epsilon\ll1$ and $r_{0}\approx\mathcal{O}(1)$. Since, as explained earlier, our reservoir state size is small, $~\mathcal{O}(\epsilon)$, in this displaced frame so to good approximation we can decompose our full density matrix with both storage modes and the reservoir ($\rho_{abr}$) as follows, where $\rho_{ij}$ signifies the reduced density matrix element of the storage cavities alone entangled to the reservoir state $\ket{i}\bra{j}$:
\begin{align}
    \rho_{abr} = &\rho_{00}\ket{0}\bra{0} + \epsilon(\rho_{01}\ket{0}\bra{1}+\rho_{10}\ket{1}\bra{0})\\\nonumber &+\epsilon^2(\rho_{11}\ket{1}\bra{1}+\rho_{02}\ket{0}\bra{2}+\rho_{20}\ket{2}\bra{0})
\end{align}
We will now take $\Delta_d=(\Delta_a-K_{ar}|r_0|^2)=(\Delta_b-K_{br}|r_0|^2)=0$ because experimentally we drive the reservoir on resonance after the Stark shift is accounted for and we have calibrated our drives such that $2\Delta_{sd} = K_{ar}|r_0|^2+K_{br}|r_0|^2+K_{ar}|\xi_{p}|^{2}+K_{br}|\xi_{p}|^{2}$.  For a $\delta = 0$ state, this calibrated equality is analogous to setting $(\Delta_a-K_{ar}|r_0|^2)=(\Delta_b-K_{br}|r_0|^2)=0$. This brings us to the following master equation for our density matrix $\rho_{abr}$:
\begin{align}
\frac{d}{dt}\rho_{abr} = -i[\hat{H}',\rho_{abr}] + \kappa_r D[\hat{r}]\rho_{abr} \\\nonumber+ \kappa_a D[\hat{a}]\rho_{abr} +
\kappa_b D[\hat{b}]\rho_{abr}
\end{align}

We can now use this to derive the dynamics of just the storage cavities ($\rho$) from getting $\mathrm{Tr}_{r}[\rho_{abr}] = \rho_{00}+\epsilon^{2}\rho_{11}$ up to second order in $\epsilon$. First, looking at the evolution of $\rho_{00}$ by multiplying our master equation by $\bra{0}$ and $\ket{0}$, we get:
\begin{align}
    \frac{1}{\kappa_{r}}\frac{d}{dt}\rho_{00} = &  -\frac{i}{\kappa_{r}}\bra{0}[\hat{H}',\rho_{abr}]\ket{0} + \epsilon^{2}\rho_{11} \\\nonumber 
    & +  \frac{\kappa_a}{\kappa_r}D[\hat{a}]\rho_{abr} +
    \frac{\kappa_b}{\kappa_r}D[\hat{b}]\rho_{abr} \\\nonumber
     = & -i\epsilon^2(\hat{A}^{\dagger}\rho_{10}-\rho_{01}\hat{A}) \\\nonumber
     & -i[-\frac{K_{aa}}{2\kappa_r}(\hat{a}^{\dagger}\hat{a})^{2}-\frac{K_{bb}}{2\kappa_r}(\hat{b}^{\dagger}\hat{b})^{2} - \frac{K_{ab}}{\kappa_r}\hat{a}^{\dagger}\hat{a}\hat{b}^{\dagger}\hat{b} \\\nonumber
     & +\frac{g_{ab}}{\kappa_r}\hat{a}^{\dagger}\hat{b}^{\dagger}(r_{0})  + \frac{g_{ab}}{\kappa_r}^*\hat{a}\hat{b}(r_{0}^{*}),\rho_{00}] \\\nonumber 
     & + \epsilon^{2}\rho_{11}+ 
    \frac{\kappa_a}{\kappa_r}D[\hat{a}]\rho_{abr} +
    \frac{\kappa_b}{\kappa_r}D[\hat{b}]\rho_{abr} 
\end{align}
where $A\equiv \frac{1}{\epsilon\kappa_r}(g_{ab}^{*}\hat{a}\hat{b}-K_{ar}r_{0}\hat{a}^{\dagger}\hat{a}-K_{br}r_{0}\hat{b}^{\dagger}\hat{b})$ thus making $||A|| \approx \mathcal{O}(1)$. Now we can find similar evolution equations for $\rho_{10}$ and $\rho_{11}$, working only up to zeroth order in $\epsilon$:
\begin{align}
    \frac{1}{\kappa_{r}}\frac{d}{dt}\rho_{10} & = -i\hat{A}\rho_{00} - \frac{1}{2}\rho_{10} +\mathcal{O}(\epsilon)
    \\\nonumber
    \frac{1}{\kappa_{r}}\frac{d}{dt}\rho_{11} & = -i(\hat{A}\rho_{01}-\rho_{01}\hat{A}^{\dagger})-\rho_{11} +\mathcal{O}(\epsilon)
\end{align}
Looking at the the equation for $\rho_{10}$, we can see that the first term involving $\rho_{00}$ is time dependant, making an exact solution difficult, but the time rate of change can be seen to be slow, on the order of $\epsilon$, whereas the second term, which acts as a damping term, is of order unity. Since the effective driving term is slowly changing and the damping term is relatively much stronger, we can impose an adiabatic elimination and take $\rho_{10}$ to be in it's steady state. The same argument would apply to $\rho_{11}$, yielding:
\begin{align}
    \rho_{10} & = -2i\hat{A}\rho_{00} + \mathcal{O}(\epsilon) \\\nonumber
    \rho_{11} & = -i(\hat{A}\rho_{01}-\rho_{10}\hat{A}^{\dagger}) + \mathcal{O}(\epsilon) 
    = 4\hat{A}\rho_{00}\hat{A}^{\dagger} + \mathcal{O}(\epsilon)
\end{align}
substituting the above expressions for $\rho_{00}$,$\rho_{10}$ and$\rho_{01}$ into eq.~B25 yields the following master equation:
\begin{align}
    \label{eq:rho_full_supp}
    \frac{d}{dt}\rho = & -i[\hat{H}_{red},\rho] + \kappa_a D[\hat{a}]\rho +
    \kappa_b D[\hat{b}]\rho \\\nonumber 
    & + D[\frac{2 g_{ab}^{*}}{\sqrt{\kappa_{r}}}\hat{a}\hat{b} - \frac{2K_{ar}r_0}{\sqrt{\kappa_r}}\hat{a}^{\dagger}\hat{a} - \frac{2K_{br}r_0}{\sqrt{\kappa_r}}\hat{b}^{\dagger}\hat{b}]\rho
\end{align}
where
\begin{align}
    \label{eq:Hred}
    \hat{H}_{red}  = &  -\frac{K_{aa}}{2}(\hat{a}^{\dagger}\hat{a})^{2}-\frac{K_{bb}}{2}(\hat{b}^{\dagger}\hat{b})^{2} \\\nonumber
    & - K_{ab}\hat{a}^{\dagger}\hat{a}\hat{b}^{\dagger}\hat{b} + \epsilon_{ab}\hat{a}^{\dagger}\hat{b}^{\dagger} + \epsilon_{ab}^*\hat{a}\hat{b} 
\end{align}
\begin{align} 
    \epsilon_{ab} =  g_{ab}r_{0}, \;\; \kappa_{ab} =  \frac{4|g_{ab}|^{2}}{\kappa_r}
\end{align}

\subsection{Ideal steady state PCS without reservoir cross Kerr}

In this section we show that the ideal pair-photon driven dissipation process, in the absence of spurious reservoir cross Kerr and single photon loss, stabilizes a pair-coherent state as discussed in Table I.   



Starting from Eq.~(\ref{eq:rho_full_supp}) and Eq.~(\ref{eq:Hred}), this ideal scenario results in equation of motion:
\begin{equation}
    \Dot{\rho} = -i(\hat{H}_{ideal}\rho - \rho \hat{H}_{ideal}^\dagger) + \kappa_{ab} \hat{a}\hat{b} \rho \hat{a}^\dagger\hat{b}^\dagger
\end{equation}
\begin{equation}
    \hat{H}_{ideal} = \epsilon_{ab}^* \hat{a}\hat{b} + \epsilon_{ab} \hat{a}^\dagger\hat{b}^\dagger + K_\mathrm{eff}\hat{a}^\dagger\hat{b}^\dagger \hat{a}\hat{b} - i \frac{\kappa_{ab}}{2}\hat{a}^\dagger\hat{b}^\dagger \hat{a}\hat{b}
\end{equation}
If the system is in a pair coherent state $\ket{\gamma, \delta}$, we can write the instantaneous density matrix time evolution as:
\begin{align}
    \Dot{\rho} =&(\epsilon_{ab}^* \gamma - \epsilon_{ab} \gamma^* + i \kappa_{ab} |\gamma|^2)\rho \\\nonumber &+ (\epsilon_{ab} + K_\mathrm{eff}\gamma - i\frac{\kappa_{ab}}{2} \gamma)\hat{a}^\dagger\hat{b}^\dagger \rho \\\nonumber &- (\epsilon_{ab}^* + K_\mathrm{eff}\gamma^* + i\frac{\kappa_{ab}}{2} \gamma^*) \rho \hat{a}\hat{b}
\end{align}
Setting the terms that distort $\rho$ to 0 to give a steady-state solution and the following expression for the PCS state size $\gamma$:
\begin{equation}
    \gamma = \frac{\epsilon_{ab}}{\frac{i\kappa_{ab}}{2}-K_\mathrm{eff}}
\end{equation}

\section{Numerical simulation methods}\label{Simulations}

We numerically simulate the PCS stabilization with two-photon pumping and dissipation that compares well with the experiment. The data shown in Fig.~\ref{fig:PC_2}(d) is a measurement of storage cavity population while pumping the PCS for variable time and the inset is measuring the population when only the FWM drive is on causing two-photon dissipation starting in a PCS. The Hamiltonian used in the simulation to fit the two-photon pumping and dissipation is unitary transformed so that the PCS is stationary in the drive frame. The reservoir mode is adiabatically eliminated resulting in a reduced Hamiltonian used in master equation Eq.~\ref{eq:rho_full} in the main text. We rewrite the master equation with the single photon loss operators for both the storage cavities, 
\begin{align}
     \partial_t \rho = -&i [\hat{H}_{sk} + (\epsilon_{ab} \hat{a}^{\dagger}\hat{b}^{\dagger} + h.c.), \rho] \nonumber\\ &+  \mathcal{D}  [\sqrt{\kappa_{ab}}\hat{a}\hat{b}+\zeta_{a}\hat{a}^{\dagger}\hat{a}  +\zeta_{b}\hat{b}^{\dagger}\hat{b}]\rho \nonumber\\ &+
      \mathcal{D}  [\sqrt{\kappa_{a}}\hat{a}]\rho +
      \mathcal{D}[\sqrt{\kappa_{b}}\hat{b}]\rho
\label{eq:SimME}
\end{align}
where $\hat{H}_{sk}$ is defined in Eq.~(\ref{eq:H_SK}) and, $$\zeta_{a}=-\frac{2K_{ar}r_{0}}{\sqrt{\kappa_{r}}}, \zeta_{b}=-\frac{2K_{br}r_{0}}{\sqrt{\kappa_{r}}}.$$ The storage cavity single photon loss rates, $\kappa_{a}=1/T_{1a}$, $\kappa_{b}=1/T_{1b}$, Kerr and cross-Kerr values measured during pumping are listed in Table. \ref{table:parameters}. The complex parameter $r_{0}=\frac{2\epsilon_{d}}{i\kappa_{r}-2\Delta_{d}}$, first defined in Eq.~(\ref{eq:r0_def}), is the amplitude of the reservoir displacement. We displace the reservoir with $\epsilon_{d}/2\pi=780$ kHz at a detuning $\Delta_{d}/2\pi=-185$ kHz which with a $\kappa_{r}/2\pi=780$ kHz results in a $|r_{0}|=1.8$. 

\begin{figure*}[tbp]
    \centering
    \includegraphics[scale=0.75]{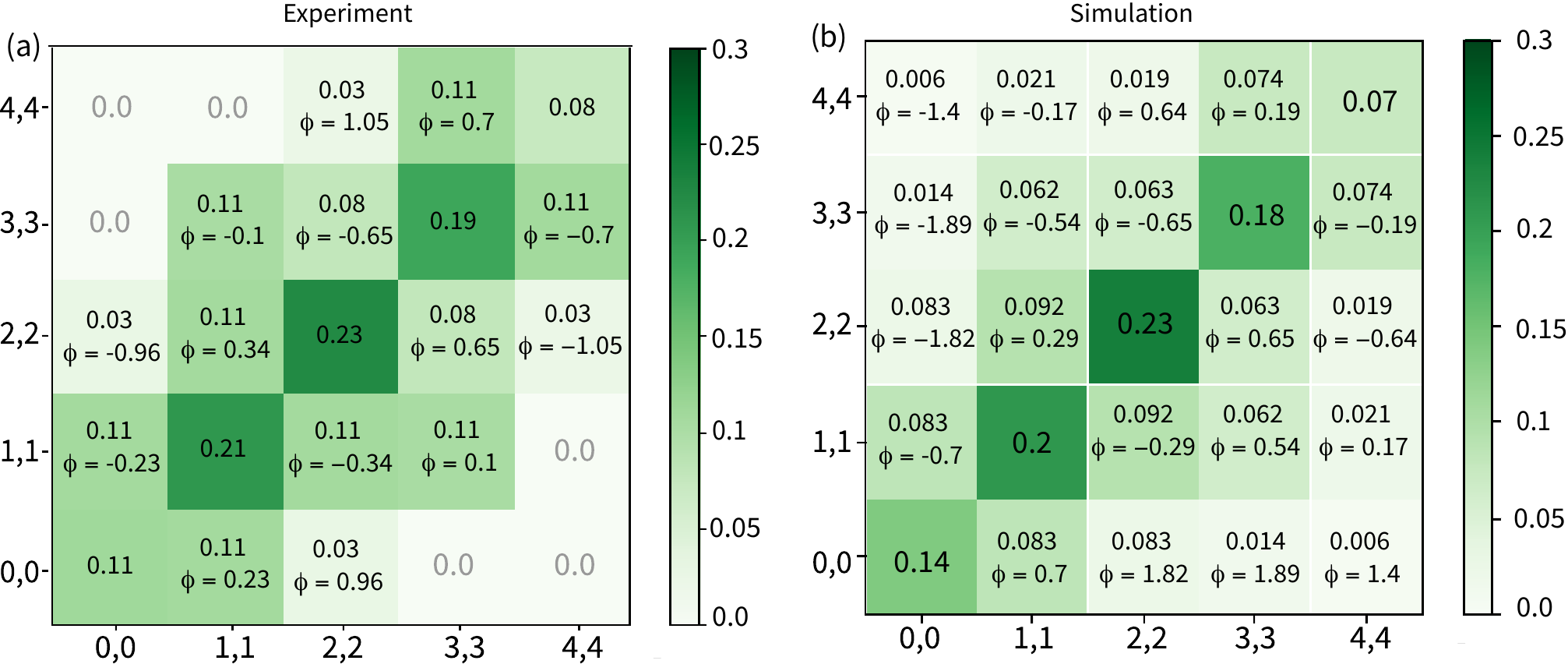}
    \caption{Density Matrix from experiment vs simulation. (a) The experimental density matrix measured using the quantum subspace tomography method discuss in the main text for $d=2$ after pumping for 15 $\mu$s and a wait time of $t_{w}=4.5$ $\mu$s. (b) The simulated density matrix with the system parameters and rates extracted from the time domain measurement fittings as in Fig.~\ref{fig:PC_2}. This is also after a 15 $\mu$s of pumping and a wait time of 4.5 $\mu$s.
}\label{fig:density_mat}
\end{figure*}

Starting in a PCS $\ket{\gamma, \delta=0}$ as shown in Fig.~\ref{fig:PC_2}(c) we measure the population of states at different times $t_{w}$ for which only the FWM tone is on (Fig.~\ref{fig:PC_2}(d inset)). In the absence of the reservoir drive the FWM drive should only cause two-photon dissipation $\mathcal{D}[\sqrt{\kappa_{ab}}\hat{a}\hat{b}]$. Using Eq.~(\ref{eq:SimME}) with $\kappa_{ab}$ as the only fit parameter and no drive term $\epsilon_{ab}=0$ the time scale for two-photon dissipation can be extracted. The dephasing rates $\zeta_{a}=\zeta_{b}=0$ since there are no photons in the reservoir and thus no dephasing caused by the storage-reservoir cross Kerr $K_{ar/br}$. In the second part we fix the two photon-dissipation $\kappa_{ab}/2\pi = 12.5$ kHz and vary $\epsilon_{ab}$. We estimate the dephasing rates from cavity $a$ and $b$, $|\zeta_{a}|^{2}/2=2\pi\times0.4$ kHz, $|\zeta_{b}|^{2}/2=2\pi\times28$ kHz respectively. We find the two-photon pumping rate to be $\epsilon_{ab}/2\pi=99$ kHz. The oscillations in the pumping time domain data are the result of the weak dissipation rate compared to the $K_\mathrm{eff}$. Simulation shows that with $K_\mathrm{eff}\gg\kappa_{ab}$, the damping of the oscillation from two-photon dissipation alone should be slower than we see in the experiment and the state is predicted by the equation for $\gamma$ as in Table.~\ref{table:PC_principle}. The state is smaller with more PCS like distribution. However, the modified dissipation operator to include dephasing effects of the cross-Kerr as derived in Appendix section~\ref{Derivation} produces faster damping, mimicking the experimental PCS distribution and size.

Using the equation, $\kappa_{ab}/2\pi = 4|g_{ab}|^2/\kappa_{r}=12.5$ kHz, we find $g_{ab}/2\pi=50$ kHz. The Stark shift we measure on the qubit due to the strong cw tone is $\sim4.3$ MHz, that makes the fourth order FWM term in the cosine expansion $g_{ab}/2\pi=60$ kHz. This value of $g_{ab}$ estimates the $\epsilon_{ab}/2\pi=120$ kHz and $\kappa_{ab}/2\pi=18.4$ kHz which is in agreement with the simulation to a good approximation. With the extracted two-photon dissipation and pumping rates we simulate the coherence and phases over time in a subspace of the state Fig.~\ref{fig:coh_over_time}(b,c).

\section{General Protocol of subspace tomography}\label{Protocol}

Using a three-level ancilla, we may use the following protocol for quantum subspace tomography:

\begin{figure}[tbp]
    \centering
    \includegraphics[scale=.37]{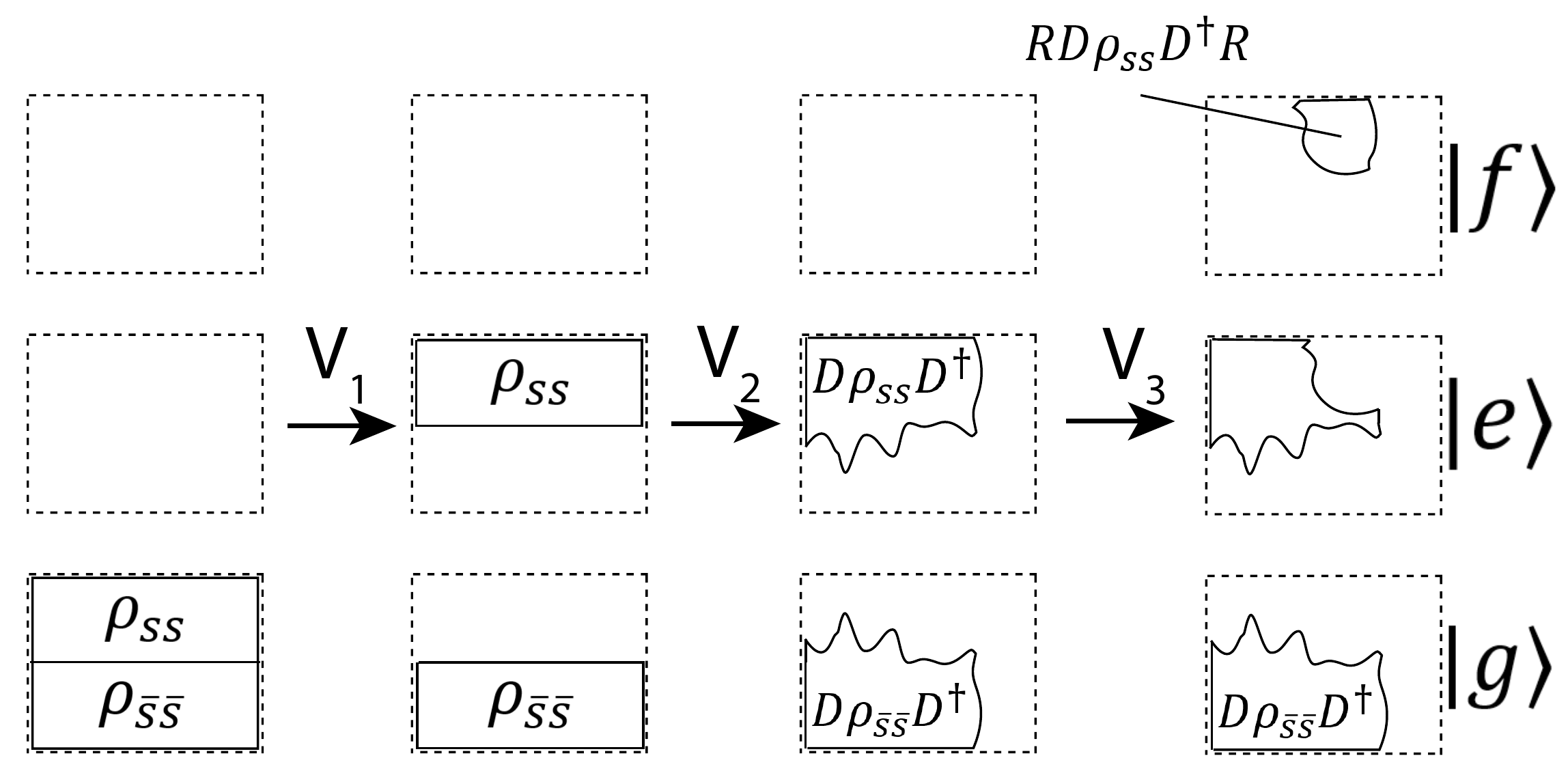}
    \caption{General protocol of subspace tomography. We are illustrating the density matrices above in a very loose manner where $\rho_{ss}$ and $\rho_{\bar{s}\bar{s}}$ are blocks of the total storage density matrix $\rho$. There are coherence elements between these two blocks that are not depicted because they do not impact the experiment at hand. The goal here is to isolate $RD\rho_{ss}D^{\dagger}R$ in the $\ket{f}$ subspace using the operations $V_{1},V_{2},V_{3}$ as shown in the image. 
    }
\end{figure}

\begin{enumerate}\addtocounter{enumi}{-1}
    \item Start with the system-ancilla state $\rho \otimes \ket{g}\bra{g}$.
    \item Perform the unitary evolution $\hat{V}_1=\exp\big[ -i\frac{\pi}{2}\hat{P}\otimes (\ket{e}\bra{g}+\ket{g}\bra{e}) \big]$, which transforms the full density matrix into $\rho_1=\hat{V}_1(\rho\otimes\ket{g}\bra{g})\hat{V}_1^\dagger=\rho_{SS}\otimes\ket{e}\bra{e}+(...)$, where the terms (...) do not contribute to $\ket{e}$ state population or later measurement. 
    
    \item Perform the unitary evolution that preserves the ancilla basis states, $\hat{V}_2=\sum_{j=g,e,f}\hat{U}_j\otimes\ket{j}\bra{j}$.  For a bosonic system, we may just apply displacement operation $\hat{U}_e=\hat{D}$, which transforms the density matrix into $\rho_2=\hat{V}_2\rho_1\hat{V}_2^\dagger=\hat{D}\rho_{SS}\hat{D}^\dagger\otimes\ket{e}\bra{e}+(...)$.
    
    \item Choose the readout projector operator $\hat{R}$ (e.g.~photon number state projection $\ket{nn}\bra{nn}$ and perform the unitary evolution $\hat{V}_3 = \exp\big[ -i\frac{\pi}{2}\hat{R}\otimes (\ket{f}\bra{e}+\ket{e}\bra{f}) \big]$.  The density matrix becomes $\rho_3=\hat{V}_3\rho_2\hat{V}_3^\dagger=\hat{R}\hat{D}\rho_{SS}\hat{D}^\dagger\hat{R}\otimes\ket{f}\bra{f}+(...)$. 
    
    \item Measure the probability in $\ket{f}$ state, Tr$[\rho_3\ket{f}\bra{f}]=\textrm{Tr}[\hat{R}\hat{D}\rho_{SS}\hat{D}^\dagger\hat{R}]$, which is sufficient for subspace tomography. 
\end{enumerate}

In principle, a measurement of the ancilla after Step 2 (post select on $\ket{e}$) can be used to physically project the cavity state to $\rho_\mathcal{SS}$ for subsequent subspace tomography.  However, this requires highly ideal measurement properties including high single-shot fidelity, quantum non-demolition on the ancilla, and no spurious back-action on the cavity system.  In comparison, in the protocol above, the only requirement on the measurement is some degree of distinguishing ability between $\ket{e}$ and $\ket{f}$.  The measurement outcome 
can be scaled from calibrated readout contrast between $\ket{e}$ and $\ket{f}$, and any spurious readout signal from $\ket{g}$ only contribute to a background independent of $\hat{D}$ and $\hat{R}$.

\section{Frequency matching in subspace tomography}\label{Freq-matching}

In a typical Ramsey-type experiment, one considers the rotating frame of a qubit or a cavity set by its first excitation pulse, and the phases of its subsequent control pulses (usally at the same frequency) at any later time can be defined relative to the first pulse and programmed in the same rotating frame.  In our subspace tomography protocol, because individual pair-wise coherence measurements involve ancilla rotation pulses of different frequencies, it is non-trivial to ensure that: 1) The subspace tomography pulse sequence carried out at any time ($t$ or $t_w$) informs the superposition phase of cavity states consistently in a pre-defined frame, and 2) The superposition phase of different Fock pairs are extracted consistently in the same rotating frame and hence can be combined in the same density matrix.  

In our experiment, we enforce a ``closed-loop" frequency-matching condition when choosing the frequency of the cavity displacement pulses ($\omega_3$ and $\omega_4$) in each of the 2d subspace tomography measurements,
\begin{equation}
	\omega_3+\omega_4 = \omega_p+\omega_d - (\omega_1-\omega_2)
\end{equation}
where $\omega_p$ and $\omega_d$ are the stabilization drive frequencies.  $\omega_1, \omega_2$ are the frequencies of the photon-number selective ancilla pulses and they must precisely match the dispersively-shifted ancilla frequencies, with the dispersive shift to be referred as $-\chi_{jk}$ for the Fock state $\ket{jk}$ of the cavities.  The closed loop in measuring the $\ket{11}$-$\ket{22}$ interference as an example is illustrated in Appendix Fig.~\ref{fig:phase_loop}. While one could in principle account for accumulated phases from the relative timing and relative frequency difference of individual pulses without this condition, this strategy allows us conveniently measure the two-cavity state in the ``pump frame" defined by the two stabilization drives self-consistently.  It is important to note that the main reason this equality is important to experimentally satisfy is that the dispersive shifts on the qubit due to different number states ($\chi_{jk}$) are not equal to the corresponding multiples of $\chi_{qa}$ and $\chi_{qb}$, i.e.~$\chi_{jk}\neq j\chi_{qa} + k\chi_{qb}$, due to higher 6th order terms in the cosine expansion of our hamiltonian that have been neglected. Due to these neglected higher order terms, $\omega_{3}$ and $\omega_{4}$ must be determined from the experimentally measured $\chi_{jk}$ values for each pair of number states to satisfy the equality.

\begin{figure}[tbp]
    \centering
    \includegraphics[scale=0.45]{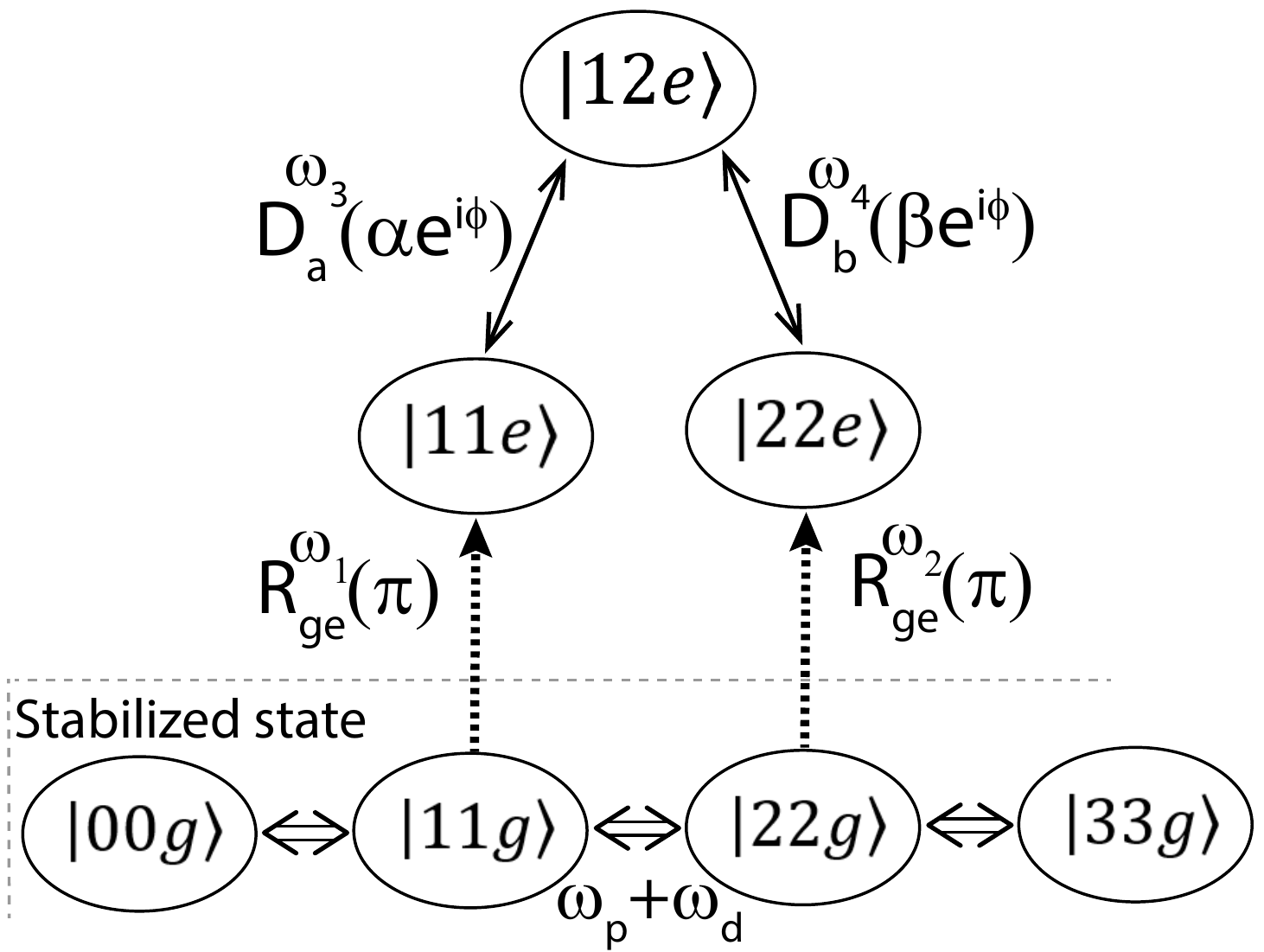}
    \caption{Visual depiction of the pulse sequence used for the subspace tomography to determine the $\rho_{11,22}$ element while tracking the $\ket{12}$ state as a specific example.
    The frequencies of the four pulses applied in this tomographic analysis and the frequencies of the original cavity stabilization drives satisfy the ``closed-loop" condition for phase locking: $\omega_1-\omega_2=\omega_p+\omega_d-\omega_{3}-\omega_{4}$ which is the same effective equality as in Eq.~(\ref{eq:simple phase match}).  Note that $\omega_1-\omega_2 = \chi_{22}-\chi_{11}\approx\chi_{qa}+\chi_{qb}$ but is not exactly equal because of the 6th-order dispersive shifts.  
    \label{fig:phase_loop}
}
\end{figure}

In the following, we show how our subspace tomography protocol faithfully extracts the superposition phase $\Phi_{jj}-\Phi_{kk}$ between two Fock components $\ket{jj}$ and $\ket{kk}$ of the cavity state as defined in the pump frame.  (The analysis is easily extensible to $\delta\neq0$ states.)  In this analysis, we work in a rotating frame where the transmon rotates with frequency $\omega_q$ and the two cavities rotate by a combined frequency $\omega_p+\omega_d$. (More specifically, Cavity $a$ rotates by $\omega_a-\Delta_{sd}/2$ and Cavity $b$ rotates by $\omega_b-\Delta_{sd}/2$ where $\Delta_{sd}=(\omega_a+\omega_b)-(\omega_p+\omega_d)$).  In this rotating frame, the frequencies of the four tomography tones, $\omega_1, \omega_2, \omega_3, \omega_4$ are on the order of the dispersive shifts, and the eigen-frequencies of the joint Fock states $\ket{jj}$ are: 
\begin{align}
\omega_{jjg} = j\Delta_{sd} - K_{ab}j^{2} - (K_{aa}+K_{bb})j(j-1)/2
\end{align}
\begin{align}
\omega_{jje} = j\Delta_{sd} - K_{ab}j^{2} - (K_{aa}+K_{bb})j(j-1)/2 - \chi_{jj}
\end{align}
for the transmon in $\ket{g}$ and $\ket{e}$, respectively. 

We now track the phase accumulation on the $\ket{jj}$ component of the cavity state over the 3 time steps of a subspace tomography, first, the wait time $(t_{w}$ a few $\mu$s after the PCS is made, next, the duration of our selective qubit pulse $(\delta t_{q}=2$ $\mu$s) that excites it to be entangled with $\ket{e}$ and finally, the duration of our displacement cavity pulses ($\delta t_{d}=24$ ns) to create interference of $\ket{jje}$ versus a different cavity state. The short duration and therefore the broad selectivity of our displacement cavity pulses gives us the freedom to vary their frequencies to satisfy this closed loop condition without needing to worry about their frequencies being too far from the desired cavity states to be displaced. 
During the first wait time period ($t_{w}$) the qubit remains in $\ket{g}$ thus accumulating a phase $\omega_{jjg}t_{w}$. During the qubit pulse step ($\delta t_{q}$) the generator for the qubit pulse is rotating in a different frame with a frequency offset of $-\chi_{jj}$ thus leading to a phase offset of $-\chi_{jj} (t_{w}+\delta t_{q})$ from the phase imparted by the generator since the generator would have been accumulating this phase through both time periods $t_{w}$ and $\delta t_{q}$. We can use the same time steps here for different generators in different frames because they all share a common reference of how t=0 is defined, where any reference point of t=0 is valid provided it is consistent among all generators. For our final time period we have a state phase accumulation from the state with the qubit in $\ket{e}$ giving $\omega_{jje}\delta t_{d}$ in addition to the phase imparted by our displacement pulse. The displacement pulses are in different rotating frames with a combined frequency of $\omega_{3}+\omega_{4}$. Taking the state that we are interfering with as $\ket{kk}$, this will impart a phase of $(\omega_{3}+\omega_{4})|k-j|(t_{w} + \delta t_{q} + \delta t_{d})$, where the $|k-j|$ factor comes from the fact that the phase gets imparted for each excitation traded. With this, we can subtract the accumulated phases between $\ket{jj}$ and $\ket{kk}$ over the total time $t_{tot} = t_{w} + \delta t_{q} + \delta t_{d}$ where we are tracking state $\ket{jj}$ so that is the only one that will recieve the $\omega_{3}+\omega_{4}$ term. The chosen state to track here will not matter so this is still a general treatment, with the phase subtraction being:
\begin{align} \label{eq:phase match}
    &\Phi_{jj} - \Phi_{kk} + (\omega_{jje} - \omega_{kke} + (\omega_{3}+\omega_{4})|k-j|)t_{tot}\nonumber \\
    = &\Phi_{jj} - \Phi_{kk} + (\omega_{jjg} - \omega_{kkg} + \nonumber\\&(\chi_{kk}- \chi_{jj} + |k-j|(\omega_{3}+\omega_{4}))t_{tot}
\end{align}
It is clear that we can simply calibrate our $\omega_{3},\omega_{4}$ for each set of interference measurements on states $\ket{jj},\ket{kk}$ to satisfy the equality:
\begin{align} \label{eq:simple phase match}
    \chi_{jj}- \chi_{kk}=|k-j|(\omega_{3}+\omega_{4})
\end{align}
so we can neglect a large part of Eq.~(\ref{eq:phase match}), leaving only $\Phi_{jj} - \Phi_{kk} + (\omega_{jjg} - \omega_{kkg})t_{tot}$.  It is important to note that this cancellation we have imparted gets rid of any terms that could have contributions from higher order neglected terms mentioned earlier and also allows an arbitrary choice of a consistent $t=0$ definition across generators. 
This leaves us with the simple linear phase accumulation terms $\omega_{jjg},\omega_{kkg}$ that can be simply calculated and subtracted away, leaving the desired phase difference of $\Phi_{jj}-\Phi_{kk}$. This linear phase accumulation from $\omega_{jjg}-\omega_{kkg}$ can be seen for different states in Fig.~\ref{fig:coh_over_time}(a) at different $t_{w}$ values, allowing one to linearly extrapolate the original phase from the $t_{tot}$ times. 


One can think of this method in a more direct intuitive way by looking at the case of isolating $\ket{11}$ and $\ket{22}$ to determine the coherence element $\rho_{11,22}$ and seeing that on top of the desired $\Phi_{22}-\Phi_{11}$ and $(\omega_{22g}-\omega_{11g})t_{tot}$ phase contributions there will be a constant phase accumulation over time with frequency of $-\chi_{22}$ and $-\chi_{11}$ on the $\ket{11}$ and $\ket{22}$ states, respectively. This is because while the cavity state $\ket{11}(\ket{22})$ is entangled with the qubit in $\ket{g}$ the generator that will send the pulse out to flip the state is rotating with a frequency of $-\chi_{11}(-\chi_{22})$, thus imparting the accumulated phase from this frequency difference and after the rotation, there will be the same frequency difference from the fact that the qubit is now in $\ket{e}$ so it will also rotate at a frequency of $-\chi_{11}(-\chi_{22})$ in this frame so in effect this frequency accumulation will always be present. As mentioned earlier, there will additionally be the phase imparted from the displacement pulse, so one simply needs to tune the $\omega_{3}, \omega_{4}$ values to satisfy the above inequality in Eq.~(\ref{eq:simple phase match}) to cancel out the effects from the $-\chi_{11},-\chi_{22}$ terms, thus leaving only the desired phase difference of $\Phi_{22}-\Phi_{11}$ and linear phase accumulation from $\omega_{22g}-\omega_{11g}$ that can be easily subtracted away.

\end{appendix}

\bibliography{Zotero_Jeff}

\end{document}